\documentclass[fleqn,usenatbib]{mnras}

\usepackage{newtxtext,newtxmath}
\usepackage[T1]{fontenc}
\usepackage{ae,aecompl}
\usepackage{adjustbox}
\usepackage{amsfonts}
\usepackage{amsmath}
\usepackage{graphicx}	
\usepackage{tabularx}
\usepackage{verbatim}
\usepackage{xspace}
\usepackage{adjustbox}
\usepackage{rotating, lipsum, babel}

\newcommand{\pc}{\,\mathrm{pc}}
\newcommand{\Msun}{\,\mathrm{M}_{\odot}}
\newcommand{\kpc}{\,\mathrm{kpc}}
\newcommand{\Gyr}{\,\mathrm{Gyr}}
\newcommand{\Gyrs}{\,\mathrm{Gyrs}}
\newcommand{\kms}{\,\mathrm{km\,s}^{-1}}

\newcommand{\Rtidal}{R_\mathrm{tidal}}
\newcommand{\rmax}{r_\mathrm{max}}
\newcommand{\rmin}{r_\mathrm{min}}
\newcommand{\Rapo}{R_\mathrm{apo}}
\newcommand{\Rpre}{R_\mathrm{pre}}
\newcommand{\fesc}{f_\mathrm{esc}}
\newcommand{\Tesc}{T_\mathrm{esc}}

\newcommand{\Mv}{M_\mathrm{v}}

\newcommand{\rs}{r_\mathrm{s}}
\newcommand{\rh}{r_\mathrm{h}}
\newcommand{\RG}{R_\mathrm{G}}
\newcommand{\Rc}{R_\mathrm{c}}

\newcommand{\Mb}{M_\mathrm{b}}
\newcommand{\Md}{M_\mathrm{d}}
\newcommand{\Mtotal}{M_\mathrm{total}}

\newcommand{\phib}{\phi_\mathrm{bulge}}
\newcommand{\phid}{\phi_\mathrm{disc}}
\newcommand{\phih}{\phi_\mathrm{halo}}
\newcommand{\phiMW}{\phi_\mathrm{MW}}

\newcommand{\Vinfty}{V_\infty}
\newcommand{\Vesc}{V_\mathrm{esc}}
\newcommand{\VLOS}{V_\mathrm{LOS}}

\newcommand{\Nbody}{$N$-body\xspace}

\DeclareMathOperator{\sech}{sech}

\newcommand{\secref}[1]{Section~\ref{#1}}
\newcommand{\figref}[1]{Figure~\ref{#1}}
\newcommand{\tabref}[1]{Table~\ref{#1}}
\newcommand{\equref}[1]{Equation~\eqref{#1}}

\title[The Escape of GCs from the dSphs of the MW]{The Escape of Globular Clusters from the Satellite Dwarf Galaxies of the Milky Way}

\author[Rostami et al.]{
Ali Rostami Shirazi,$^{1}$\thanks{E-mail: a.rostami@iasbs.ac.ir}
Hosein Haghi,$^{1}$
Pouria Khalaj,$^{1}$
Ahmad Farhani Asl$^{1}$ and \newauthor
Akram Hasani Zonoozi$^{1,2}$
\\
$^{1}$Department of Physics, Institute for Advanced Studies in Basic Sciences (IASBS), PO Box 11365-9161, Zanjan, Iran\\
$^{2}$Helmholtz-Institut f\"ur Strahlen-und Kernphysik (HISKP), Universit\"at Bonn, Nussallee 14-16, D-53115 Bonn, Germany\\
}

\date{Accepted XXX. Received YYY; in original form ZZZ}

\pubyear{2021}

\begin{document}
\label{firstpage}
\pagerange{\pageref{firstpage}--\pageref{lastpage}}
\maketitle

%%%%%%%%%%%%%%%%%%%%%%%%%%%%%%%%%%%%%%%%%%%%%%%%%%
\begin{abstract}
 Using numerical simulations, we have studied the escape of globular clusters (GCs) from the satellite dwarf spheroidal galaxies (dSphs) of the Milky Way (MW). We start by following the orbits of a large sample of GCs around dSphs in the presence of the MW potential field. We then obtain the fraction of GCs leaving their host dSphs within a Hubble Time. We model dSphs by a Hernquist density profile with masses between $10^7\Msun$ and $7\times 10^9\Msun$. All dSphs lie on the Galactic disc plane, but they have different orbital eccentricities and apogalactic distances. We compute the escape fraction of GCs from 13 of the most massive dSphs of the MW, using their realistic orbits around the MW (as determined by \textit{Gaia}). The escape fraction of GCs from 13 dSphs is in the range  $12\%$ to $93\%$. The average escape time of GCs from these dSphs was less than 8 $\Gyrs$, indicating that the escape process of GCs from dSphs was over. We then adopt a set of observationally-constrained density profiles for specific case of the Fornax dSph. According to our results, the escape fraction of GCs shows a negative correlation with both the mass and the apogalactic distance of the dSphs, as well as a positive correlation with the orbital eccentricity of dSphs.  In particular, we find that the escape fraction of GCs from the Fornax dSph is between $13\%$ and $38\%$. Finally, we observe that when GCs leave their host dSphs, their final orbit around the MW does not differ much from their host dSphs. 

\end{abstract}


%%%%%%%%%%%%%%%%%%%%%%%%%%%%%%%%%%%%%%%%%%%%%%%%%%%%%%%%%%%%%%%%%%%%%%%%%%%%%
\section{Introduction}
Cosmological simulations suggest that galaxies may form through two consecutive formation phases. Initially, star formation occurs within the parent galaxy and subsequently, significant fraction of the final mass is accreted \citep{Oser2010,Cook2016,Rodriguez2016}. Moreover, many studies show that clustered star formation is dominant over isolated star formation and that star clusters constitute the fundamental building blocks of galaxies \citep{Lada2003, Kroupa2005, Parker2007, Baumgardt2008, Assmann2011, Parker2014}. Therefore, one can expect massive galaxies to contain globular clusters (GCs) as well as field stars formed both ex-situ (accreted from satellite galaxies) and in-situ (see e.g. \citealt{Pillepich2015}).

\par More than 50 dwarf galaxies and about 150 GCs are currently known in the Milky Way (MW) galaxy \citep{Helmi2018, Baumgardt2019, Vasiliev2019, Massari2019, Drlica-Wagner2020}. The exact formation mechanism of GCs is still debated \citep{Brodie2006, Forbes2018}. Nevertheless, one can utilise their properties such as age, metallicities and orbital parameters to differentiate GCs formed in-situ and accreted GCs from dwarf satellite galaxies (e.g. \citealt{Cote1999}). The in-situ formed GCs can shed light on the star formation history and the chemical enrichment of their host galaxy \citep{Pfeffer2018, Kruijssen2020}. Accreted GCs, on the other hand, are among the few observable fossil records of the hierarchical merging process which yielded the stellar halo of the host galaxy \citep{Forbes2018, Forbes2020}. Therefore, discerning between these two types of GC populations within a galaxy, can aid us to unveil the history of accretion and baryionic processes in galaxies (e.g. \citealt{Bell2017, Oman2017, Youakim2020}).

\par There are evidences that the MW has been probably built up from mergers with smaller dwarf galaxies \citep{Searle1978} and GCs can be considered as potential tracers of this process. If a dwarf galaxy merges with the MW its GCs are expected to be incorporated into the MW’s GC population. Some GCs in our galaxy, especially the so-called `young halo' population, are thought to have been accreted from dwarf galaxies \citep{Law2010,Mackey2004}. Direct evidence for this scenario is provided by the Sagittarius (Sgr) dwarf spheroidal galaxy (dSph) that is in the course of merging with the MW \citep{Ibata1994}. During the last 7 billion years, the Sgr dSph has undergone repeated encounters (during close pericentric passages) with the MW inducing episodes of enhanced star formation in the Galaxy \citep{Ruiz-Lara2020}. Intriguingly, one of these episodes occurred at about $5-6$ billion years ago \citep{Ruiz-Lara2020}. This approximately matches the age of the Solar System, which is estimated to be $\approx 4.6\Gyr$ \citep{Bouvier2010, Bonanno2015}, indicating that such intergalactic interactions can have profound effects even down to the scale of planetary systems. There is also a growing consensus that a significant fraction of the MW GCs are accreted systems, while the rest are formed in the early phase of galaxy formation \citep{Forbes2010}. For example, a comparison between the MW GCs and those in satellite galaxies shows that all 30 of the young halo GCs, which exhibit a more extended distribution, as well as $10-12$ old halo GCs may have originated in dwarf galaxies. This observation hints that the MW has merged with approximately 7 dSphs \citep{Mackey2004}, reaffirming that Galactic GCs are formed both in-situ and in dwarf galaxies later accreted on to the MW. Further evidence comes from GCs such as Pal 4 and Pal 14 which reside in the outskirt of the MW halo. These GCs exhibit a shallow mass function slope and a high degree of mass segregation \citep{Jordi2010,Frank2013}, which cannot merely be explained by two-body relaxation, if one assumes a canonical \citet{Kroupa2001} initial mass function for the GCs. However, a solution to this discrepancy, as suggested by the outcome of \Nbody simulations, could be that these GCs formed in dwarf satellite galaxies and was later accreted by the MW \citep{Zonoozi2011,Zonoozi2014,Zonoozi2017}. 

\par Moreover, most of massive galaxies evince a bimodality in the colour distribution of their GC systems, corresponding to two sub-populations of blue and red clusters \citep{Larsen2001,Peng2006}. Spectroscopic studies ascribe the observed colour bimodality to a metallicity difference between the two populations \citep{Zinn1985, Armandroff1988}. The redder and more metal-rich globular clusters formed in situ, while the bluer and probably more metal-poor GCs may have been accreted from smaller satellite dwarf galaxies \citep{Brodie2006, Cote1998}. In addition, the phase-space data indicates that GCs with disc-like orbits (i.e. the in-situ phase of galaxy formation) occupy the more metal-rich arm, while the more metal-poor GCs are associated with the accreted halo GCs. The age and metallicity of a significant number of the Galactic GCs are incompatible with that of the MW, implying that the MW is unlikely to be the origin of such GCs. The observed `bifurcated' age-metallicity relation and kinematic information allow for the separation of GCs into accreted vs. in-situ \citep{Forbes2010, Marin2009, Leaman2013a}. In addition, the observed age spreads and metallicity offset in the metal-poor sequence of dwarf galaxies \citep{Leaman2013b} require an accretion origin. As an example, there are indications that the origin of Canis Major GCs are different from that of MW GCs on the basis of different parameters such as age, metallicity, galactocentric distance, magnitude, and half-mass radius \citep{Forbes2004}.

\par Only four of the MW companions, namely the Large Magellanic Cloud (LMC), the Small Magellanic Cloud (SMC), the Fornax dSph, and the Sgr dSph contain well-known GCs. LMC, Fornax, and Sgr have GCs that are as old as the oldest Galactic GCs \citep{Grebel2004}. Deep colour-magnitude diagrams obtained with the Hubble Space Telescope, \citet{Glatt2008} showed that a GC in the SMC (NGC 121) is approximately $2-3\Gyrs$ younger than the oldest Galactic GCs and resembles the young halo metal-poor GCs. It is not clear why the SMC formed this GC a few $\Gyrs$ later than other nearby dwarf galaxies containing GCs.

\par \cite{Martin2004} confirmed the existence of a disrupted dwarf galaxy, i.e. the `Monoceros Ring' discovered by \cite{Newberg2002}. By comparing the phase-space distribution of Galactic GCs with M-giant stars, \cite{Crane2003} and \cite{Frinchaboy2004} have shown that NGC 2298, NGC 2808, NGC 5286, Pal 1, and BH 176 (all which are GCs of the MW) are most likely accreted from the Monoceros Ring. In addition, \citet{Danny2021} reported the discovery of a so-called `inner Galaxy structure' which is probably the remnant of a satellite galaxy accreted to the MW in its early phase of formation. 

\par In the Local Group, Fornax dSph possesses the highest GC specific frequency with $\Mv\approx-13.2$ and 6 GCs \citep{Pace2021}. Many evidence have been presented for accretion to the MW from the Sgr GCs, Terzan 7, Terzan 8, Arp 2, and M54 \citep{Ibata1995, Layden2003}. 
Based on the angular position and velocity of the Sgr dSph stream, \cite{Law2010} concluded that 5 GCs (Arp 2, M 54, NGC 5634, Terzan 8, and Whiting 1) are associated with the Sgr dSph, as well as 4 other GCs (Berkeley 29, NGC 5053, Pal 12, and Terzan 7), albeit with a lower probability. Using \textit{Gaia} Data Release 2 (DR2), \cite{Bellazzini2020} confirmed the membership of Pal 12 and Whiting 1 and nominated 3 other candidates (NGC 2419, NGC 5634 and NGC 4147) to be also possibly associated with the Sgr stream, as the clusters are consistent with the age-metallicity relation of the stars in the Sgr main body. In the case of Pal 12, the cluster shows as chemical “oddities” as Sgr with MW, thus confirming extra-galactic origin of the GC \citep{Sbordone2007}. Besides, by obtaining new estimations for the age, metallicity and the distance of Whiting 1, it seems likely that the cluster is among the 6 GCs that are strongly evidenced to be originated from Sgr \citep{Carraro2007}.

\par In the present paper, which is the first in a series, we examine the escape of GCs from dwarf satellite galaxies and their capture by the MW. We will perform numerical integrations for a large number of GCs that initially start in a dwarf galaxy and are then detached from their host due to the tidal forces exerted by the MW and finally assume an orbit in the Galactic halo. In particular, we study the effect of different orbital parameters and density models of dSphs on the escape fraction of their GCs, with a focus on 13 of the most massive dSphs of the MW, especially the Fornax dSph. The main question that we ultimately aim to answer in these series of papers, is whether the percentage of the escaped GCs within a Hubble time is consistent with the number of possible candidate GCs that is assumed to be accreted by the MW. It is also possible that some GCs have been ejected from other galaxies and are fully dissolved in the MW. However, this will not be discussed in the present paper.

\par The paper is structured as follow. In \secref{sec:methodology} we describe the methodology and the initial conditions used in our analysis. \secref{sec:results} embodies the main results including the escape fraction and the average escape time of GCs. Finally, we summarise and conclude our work in \secref{sec:conclusion}.  

%%%%%%%%%%%%%%%%%%%%%%%%%%%%%%%%%%%%%%%%%%%%%%%%%%%%%%%%%%%%%%%%%%%%%%%%%%%%%
\section{Methodology and initial conditions}\label{sec:methodology}
The two-body relaxation time of galaxies, including dwarf galaxies, is significantly longer than a Hubble time. In fact, the relaxation time can be used as a dynamical criterion to separate galaxies from star clusters \citep{Forbes2011}. As a result, one can treat dwarf galaxies as collisionless systems meaning that we do not need \Nbody simulations to reconstruct the potential field of a dSph. Instead, we can reduce the problem to a three-body system consisting of the MW galaxy, a satellite dwarf galaxy (i.e. a dSph) and a GC (represented by a point mass). The three-body problem in its general form has 18 phase-space parameters\footnote{For each object there are 6 parameters, i.e. 3 for position and 3 for momentum. As a result, for a three-body system the total number of parameters adds up to 18. This is different from the degrees of freedom of the system which is smaller, due to the conservation of energy and momentum.} that needs to be determined. In our case, the mass of a GC compared to the MW and a dSph is negligible, and the GC virtually has no effect on the motion of the other two objects. The same is true when comparing a dSph and the MW. As a result, one can further simplify the problem to a restricted three-body problem which only has 12 phase-space parameters, i.e. the MW sits still at the origin of our coordinate system. The orbit of the dSph is only determined by the MW potential, whereas the trajectory of the GC is prescribed by the combined potential field of the MW and the dSph. Compared to \Nbody simulations, this approach is computationally less expensive by several orders of magnitude which allows us to sweep a large parameter space.

\par There is not any general solution (in closed-form) to the gravitational three-body problem. Therefore, we numerically solve the equations of motion for the aforementioned three-body system, using a 10th order Runge-Kutta integrator, from $t=0\Gyr$ to $t=13.6\Gyr$ (present time). In particular, our calculations are performed assuming three different sets of initial conditions and models for dSphs. First, we obtain the escape fraction of GCs for a large ensemble of dSphs, all lying on the Galactic disc plane and sharing the same density profile, albeit with different total masses and orbital parameters (\tabref{tab:grid_parameters}). This enables us to examine how the escape fraction of GCs changes in response to different tidal fields. In the second part, we obtain the escape fraction of GCs from a number of MW dSphs using their realistic orbital parameters but a generic density model. Finally, we focus on the Fornax dSph as a realistic case study. In this case, the orbit and the density profile of Fornax are both observationally-constrained. The details of the initial conditions as well as the adopted models for dSphs and the MW are further explained below. 

%--------------------------

\subsection{The Milky Way}\label{sec:MW}
\subsubsection{Coordinate system}
We adopt a right-handed Galactocentric coordinate system whose origin lies on the centre of the Galaxy, the $x$-axis points towards the Sun, the $y$-axis points towards the disc-rotation direction, and the $z$-axis points towards the North Galactic Pole. In such a coordinate system the current position and velocity vectors of the Fornax dSph are $\vec{R_0}=(41.29\kpc, -50.95\kpc, -134.10\kpc)$ and $\vec{V_0}=(-40.45\kms, -126.24\kms, 82.96\kms$), respectively \citep{Fritz2018}.

\subsubsection{Model of the potential field}\label{sec:MW_model}
Our model for the MW potential field consists of three components, namely a central bulge, a disc, and a dark matter halo. 
\begin{equation}
    \phiMW = \phib + \phid + \phih
\end{equation}
The bulge is essentially a point mass
\begin{equation}\label{eq:phi_bulge}
	\phib  = {\frac{-G\Mb}{R}},  	
\end{equation}
where $\Mb=1.5\times 10^{10}\Msun$ is the bulge mass \citep{Xue2008} and $R=\sqrt{x^{2} + y^{2} + z^{2}}$ is the Galactocentric distance. The gravitational potential of the disc is represented by the \citet{Miyamoto1975} profile,
\begin{equation}\label{eq:phi_disc}
	\phid  = {\frac{-G\Md}{\sqrt{x^{2} + y^{2} + \left(a+\sqrt{b^{2} + z^{2} }\right)^{2}}} }
\end{equation}
where $a=4\kpc$ and $b=0.5\kpc$ are the scale length and the scale height of the disc, respectively. $\Md=5\times 10^{10}\Msun$ is the disc mass \citep{Xue2008}. Following \citet{Khalaj2016}, we adopt a logarithmic potential for the dark matter halo of the MW, 
\begin{equation}\label{eq:phi_halo}
    \phih  = {\frac{1}{2} \Vinfty^{2} \ln\left(\Rc^2 + R^2\right)},  	
\end{equation}
where $\Rc=R_{\odot}=8300\pc$ is the core radius and $\Vinfty=V(R_{\odot})=239\kms$ is the asymptotic speed of the flat rotation curve of the Galaxy \citep{McMillan2011, Irrgang2013}. 

%--------------------------
\subsection{Globular Clusters}
\subsubsection{Initial conditions}\label{sec:GCs_initial_conditions}
GCs are spatially distributed according to the density profile $(\rho)$ of their host dSph, i.e. the probability ($P$) of finding a GC within an infinitesimally thin shell of thickness $\mathrm{d}r$ located at the distance $r$ from the centre of the host dSph is
\begin{equation}\label{eq:rho_GC}
    P(r) \propto \rho(r)r^2 \mathrm{d}r
\end{equation}
The initial velocity of GCs is drawn from a three-dimensional Maxwell-Boltzmann distribution \citep{Maxwell1860A, Maxwell1860B}, assuming orientation isotropy for the velocity vector of GCs (spherical symmetry),
\begin{equation}\label{eq:maxwell}
    f(v) \propto v^2 \exp{\left(-\frac{v}{v_p}\right)^2}
\end{equation}
where $v_\mathrm{p}=\Vesc/2$ is the most probable speed and $\Vesc (r) = \sqrt{-2 \Phi}$ is the escape speed at radius $r$ from the centre of the dSph.
\par For each dSph, we follow the orbits of 450 initially-bound GCs within a Hubble time. For our purpose, this is a sufficiently large sample size which allows us to explore the entire position-momentum space, given by equations \ref{eq:rho_GC} and \ref{eq:maxwell}. To calculate the escape fraction of GCs from each dSph we need a measure to designate runaway GCs.

%--------------------------
\subsubsection{Escape criterion}\label{sec:GCs_escape_criterion}
As an escape criterion for GCs we need to have an estimation of the tidal radius of the dSph as it orbits the MW. Assuming the potential field of the MW at large radii is dominated by a dark matter halo with a logarithmic profile, the tidal radius of a dSph in such a field as derived by \citet{Khalaj2016} is
\begin{equation}\label{eq:r_titdal}
	\Rtidal = {\left(\frac{G M}{2 {\Vinfty}^2}\right)^{1/3} \RG^{2/3} \left(\left(\frac{\Rc}{\RG}\right)^2+1\right)^{2/3}},
\end{equation}
where $\RG$ is the instantaneous distance of the dSph to the Galactic centre that varies as the dSph orbits MW and $M$ is the total mass of the dSph. $\Rc$ is the core radius and $\Vinfty$ is the asymptotic speed of the Galactic rotation curve (see \secref{sec:MW_model}). 
In our analysis, we only consider the GCs which are initially bound to the dSphs. GCs that remain within $2\times\Rtidal$ for about 3 times their orbital period, are considered as initially-bound. The rest are designated as initially-unbound and therefore excluded. We follow the trajectory of initially-bound GCs for a Hubble time and mark those whose final distance from the dSph exceeds $2\times\Rtidal$. The overall escape fraction of GCs from a dSph, denoted by $\fesc$, is then obtained by averaging over the escape fractions from all radii smaller than or equal to the tidal radius of the dSph ($r\leq\Rtidal$).

%--------------------------
\begin{table}
	\centering
	\begin{tabular}{cccc}
		\hline
		Parameter & Definition & Value & $N$ \\ 
		\hline 
		$M_0$ & total mass & $[10^7-7\times10^9]\Msun$ & 9 \\ 
		$\alpha$ & scale length & $[0.1-1.0]\kpc$ & 5\\
		$e$ & orbital eccentricity & $[0.0-0.8]$ & 10 \\ $\Rapo$ & apogalactic distance & $[8.5-200]\kpc$ & 9 \\
		$i$ & orbital inclination & $0\degr$ & 1 \\
		\hline
	\end{tabular}
	\caption{The parameters range for the density profile (Hernquist model) and orbits of the dSphs. The orbital inclination is measured from the Galactic disc plane, i.e. an inclination of $90\degr$ corresponds to a polar orbit. The number of samples $(N)$ used for each parameter range is given in the last column. The total sample size is $N_\mathrm{total}=9\times5\times10\times9\times1=4050$.}
	\label{tab:grid_parameters}
\end{table}

%--------------------------
\begin{figure*}
    \centering
    \includegraphics[width=0.95\linewidth]{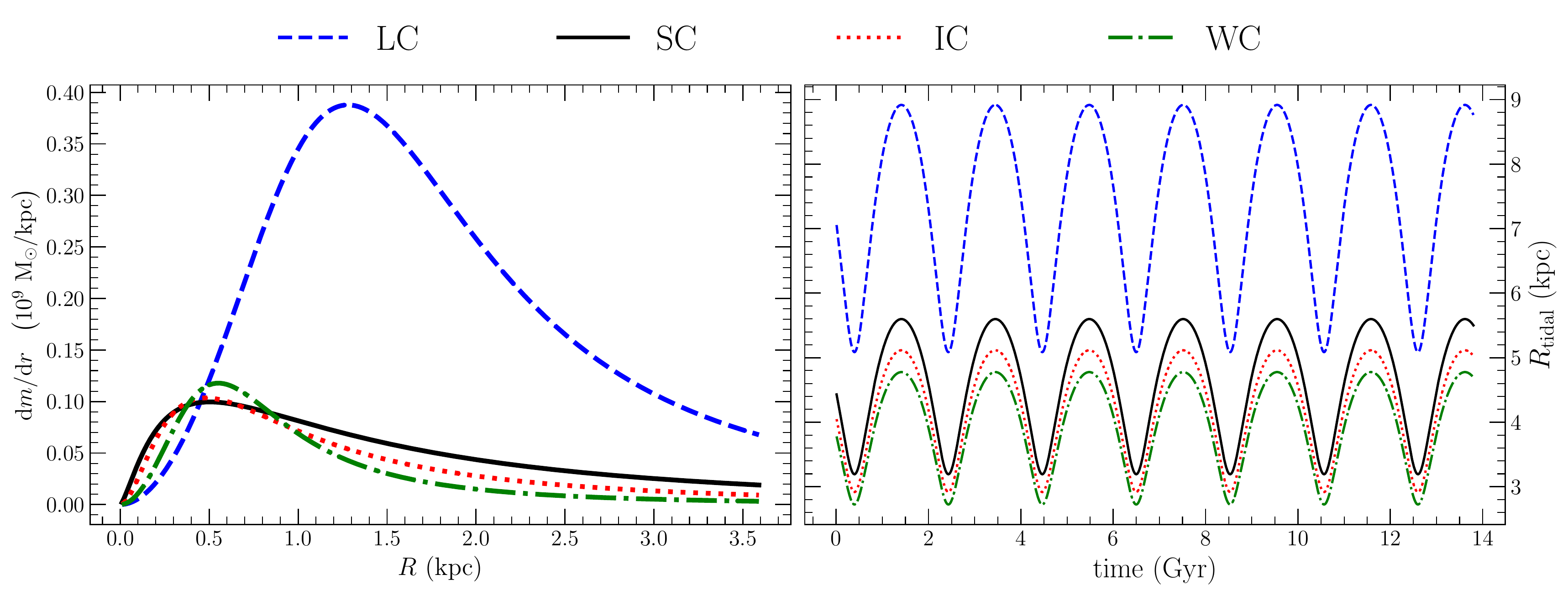}
    \caption{\textbf{Left}: Mass distribution of the Fornax dSph as a function of radius derived from four density models. The area under the curve corresponds to the total mass for each model. The LC and the WC models have the largest and the smallest total masses, respectively (see \tabref{tab:cole_params}). \textbf{Right}: The time evolution of the tidal radius of the Fornax dSph for different density models. In both panels the models are color-coded as follows: LC (dashed blue line), SC (solid black line), IC (dotted red line), WC (dash-dotted green line).}
	\label{fig:combined_profiles}
\end{figure*}

\begin{figure*}
	\centering
	\includegraphics[width=0.93\linewidth]{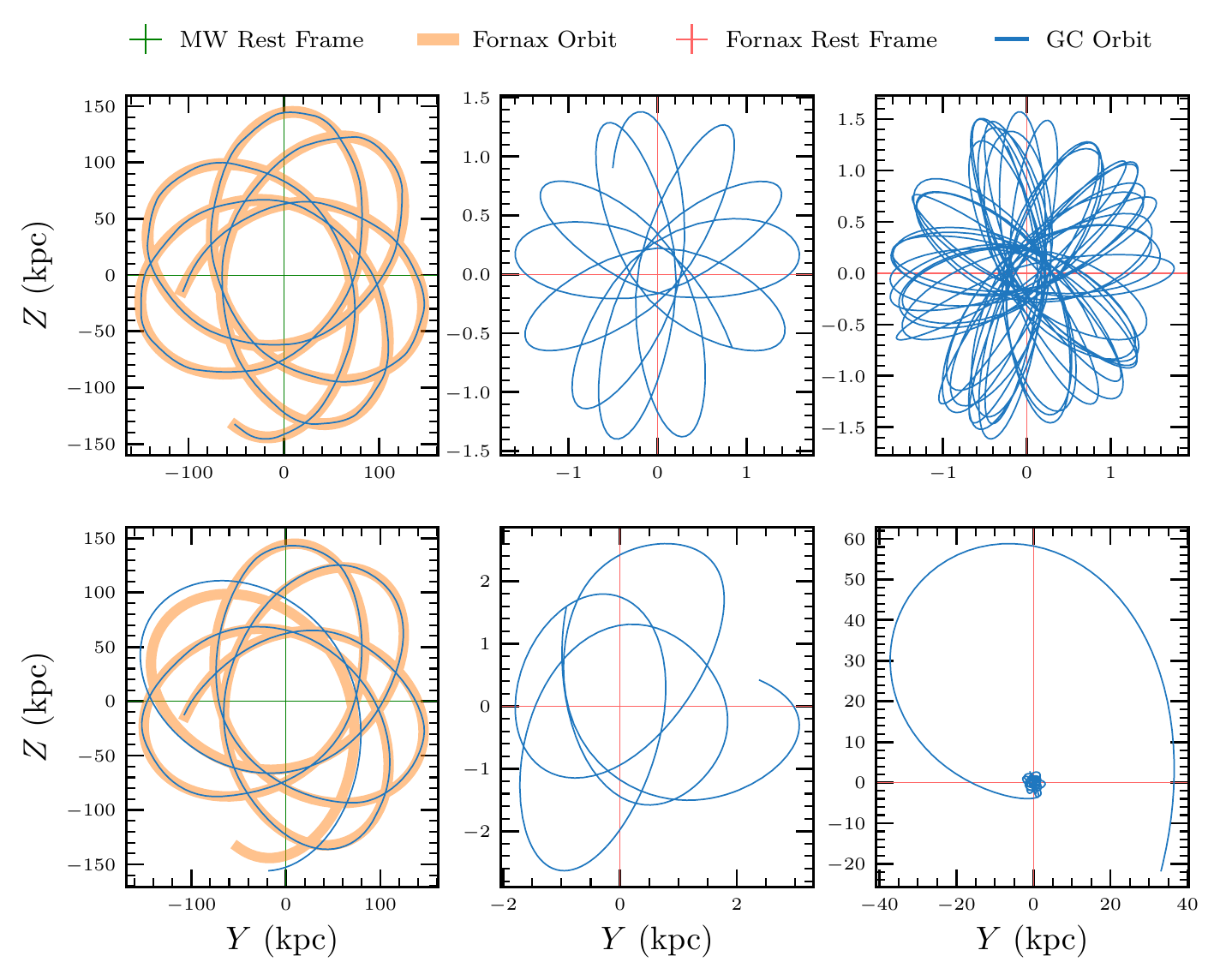}
	\caption{\textbf{Left}: The orbit of the Fornax dSph (thick orange lines) and two of its GCs (blue lines). Both GCs are initially bound. The origin marks the position of the MW. \textbf{Middle}: The orbit of the same GCs after $3\Gyrs$ which are still bound, in the non-inertial reference frame of the Fornax dSph. The centre of the Fornax lies at the origin. \textbf{Right}: As same as the middle column but after $13.6\Gyrs$. The bottom row shows a case that the GC has escaped from the Fornax dSph, whereas in the top row the GC is bound. The orbits are depicted for an instance of the SC model, and the $Y-Z$ plane of the Galactocentric coordinates.}
	\label{fig:orbits}
\end{figure*}

%--------------------------
\subsection{The sample of dSphs: the general case}\label{sec:dsph}

\subsubsection{Density profile}\label{sec:dsph_density} We assume a general model for the mass distribution within a typical dSph. This model is an isotropic \citet{Hernquist1990} profile (cf. \citealt{Baes2002}),
\begin{equation}\label{eq:rho_hernquist}
    \rho(r)=\frac{M_0}{2\pi}\frac{\alpha}{r}\frac{1}{(r+\alpha)^3}
\end{equation}
\begin{equation}\label{eq:m_hernquist}
	M(r) = \frac{M_0 r^2}{\left(r + \alpha\right)^2}
\end{equation}
where, $\alpha$ is a characteristic scale length and $M_0$ is the total mass of the dSph, ranging in $[0.1-1.0]\kpc$ and $[10^7-7\times10^9]\Msun$ respectively. The scale length $\alpha$ is related to the projected half-light radius $\rh$ via the following relation \citep{Hernquist1990},
\begin{equation}\label{eq:r_alpha}
   \rh = 1.81\alpha 
\end{equation}
This equation comes into application, when we want to convert the observed projected half-light radii of dSphs to $\alpha$, which is not an observed parameter (see \tabref{tab:MW_dsphs_params}).

\subsubsection{Initial conditions of dSphs}\label{sec:dsph_init}
A sample of dSphs on different orbits around the MW are subject to different tidal fields, which can affect the escape fraction of GCs. To examine this effect, we consider a range of different eccentricities ($e$) and apogalactic distances ($\Rapo$) for dSphs, all orbiting the MW in the Galactic disc plane ($i=0\degr$). The parameters range for $\Rapo$ and $e$ are $[8.5-200]\kpc$ and $[0.0-0.8]$, respectively. All dSphs are initially located at the apogalacticon which means the initial velocity is perpendicular to the semi-major axis and lies on the Galactic disc plane. \tabref{tab:grid_parameters} summarizes the range of the parameters defined in this section and \secref{sec:dsph_density}. We have $4050$ dSphs which adequately sample the parameter-space of $(M_0, \alpha, \Rapo, e)$. In addition, each dSph has 450 GCs (section \ref{sec:GCs_initial_conditions}). In total, we compute the orbit of $4050\times450=1822500$ GCs for a Hubble time. 

%--------------------------
\subsection{dSphs of the MW}\label{sec:MW-sat}

In addition to the aforementioned set of initial conditions for dSphs, we also obtain the escape fraction of GCs from 13 of the most massive dSphs of the MW, namely Antlia II, CanVen I, Carina I, Draco I, Fornax I, Leo I, Sculptor I, Sextans I, Ursa Major I, Ursa Minor I, SMC, LMC, and Sgr I. 

\subsubsection{Density profile}

We used the simple mass estimator of \cite{Walker2009} that relates $M(\rh)$ to observed quantities of dSphs via the following equation

\begin{equation}\label{eq:walker}
    M(\rh)=\mu \rh\sigma^2,
\end{equation}
where $M(\rh)$ is the mass within $r=\rh$, $\sigma$ is the stellar velocity dispersion, and $\mu=580\Msun\pc^{-1}\mathrm{km}^{-2}{\mathrm s}^2$. We assume that the total mass follows the light profile and hence the scale length ($\alpha$) can be obtained from the projected half-light radius (Eq. \ref{eq:r_alpha}). 
Given the mass within $\rh$, i.e. $M(\rh)$, and \equref{eq:m_hernquist} one can obtain the Hernquist parameters, namely $M_0$ and $\alpha$. These parameters for different dSphs are given in \tabref{tab:MW_dsphs} and are denoted by $M_1$ and $\alpha_1$.

Alternatively, by considering a Hernquist density profile for both dark matter halo and the baryonic component, and performing the Jeans analysis, one can obtain the velocity dispersion at different projected radii. For the baryonic component, we assume the mass to be the stellar mass \citep{Alan2012}, hereafter denoted by $M^{\prime}_{2}$. Using the projected half-light radius of dSphs and \equref{eq:r_alpha}, we obtain the scale length of the baryonic component ($\alpha^{\prime}_{2}$). Hereafter, the mass and scale length of the dark matter halo is denoted by $M_2$ and $\alpha_2$, respectively.

Fitting of the calculated velocity dispersion profile to the observed data points is achieved by adjusting the mass and the scale length of each galaxy as two free parameters in our analysis ($M_2$, $\alpha_2$). For comparison, the best-fitting model for each galaxy is shown in \figref{fig:vel_dis}. Our overall results and the values of mass and scale length for 8 dSphs are summarized in \tabref{tab:MW_dsphs_model2}. 

As it can be seen, the resulted escape fraction, $\fesc$, using the mass estimator from \equref{eq:walker} generally stands in  agreement with constraints obtained from the full Jeans analysis and it is robust against a wide range of halo models when evaluated near the half-light radius \citep{Walker2009}. Given that only the velocity dispersion profile for 8 galaxies were available, we keep the results of the first method (Model 1), which is based on \equref{eq:walker} in the main body of the paper and the results of Jeans analysis (Model 2), are presented in the Appendix. That the escape fraction is not significantly different in either case suggests that $M(<r_h)$ is more important in determining $\fesc$ than the total mass.

It should be mentioned that for all dSph models we assume that the distribution of GCs follows the distribution of baryonic matter. This is a reasonable assumption owing to the fact that more than 90 percent of GCs in the Galaxy are distributed within a radius of $10\kpc$. Such an assumption is also in agreement with \cite{Hudson2018}, who found the effective radius of the MW GC system to be $\sim4\kpc$. Later, in \secref{sec:caveat} we discuss how the results change if any other initial GC distribution is adopted.

Since, we did not find a suitable Hernquist model for SMC and LMC, we used the model proposed by \cite{bekki2009}. They adopted a Plummer model, with masses of $3\times10^9\Msun$ and $2\times10^{10} \Msun$, and scale lengths of $2\kpc$ and $3\kpc$ for SMC and LMC, respectively.

\subsubsection{Initial conditions of dSphs of the MW}\label{sec:MW_dsph_init}
We get the present-day equatorial coordinates ($\alpha, \delta$), proper motions $(\mu_\alpha\cos{\delta}, \mu_\delta)$, line-of-sight velocity ($\VLOS$), and the heliocentric distance of dSphs ($D_\odot$) from the \textit{Gaia} DR2 catalogue. We then trace back the orbit of dSphs for $13.6\Gyrs$ to obtain its initial conditions (see \tabref{tab:MW_dsphs_params}). Our calculations yield orbital eccentricity $e$, pregalactic distance $\Rpre$ and apogalactic distance $\Rapo$ of the dSphs. These realistic orbital parameters enables us to draw some intriguing comparisons explained in \secref{sec:results}. The uncertainties in line-of-sight velocity and proper motions of dSphs are taken into account in our calculations.

The MW has a massive and extended halo. This causes massive dwarfs (e.g. LMC, SMC, Sgr), which pass through the halo, to spiral inwards due to dynamical friction. For some dwarf galaxies the time scale of dynamical friction is less than $10\Gyr$, meaning that it is an important factor to consider over the lifetime of the galaxy \citep{Bar2022}. The galactocentric distance of such massive dwarfs were larger at $\sim13\Gyrs$ ago. In order to calculate the orbital history of dwarfs we adopt the following standard form of dynamical friction for the halo \citep{Binney2011}.

\begin{equation}
   F_\mathrm{DF}=-0.428\,\mathrm{ln}(\Lambda)\,\frac{GM^2}{r^2},
\end{equation}\label{eq:DYfric}
where, $r$ is the distance of the dwarf from the center of the MW, $M$ is the mass of the dwarf, and $\mathrm{ln}(\Lambda) = 3$ is the Coulomb logarithm.

%%%%%%%%%%%%%%%%%%%%%%%%%%%%%%%%%%%%%%%%%%%%%%%%%%%%%%%
%%%%%%%%%%%%%%%%%%%%%%%%%%%%%%%%%%%%%%%%%%%%%%%%%%%%%%
%%%%%%%%%%%%%%%%%%%%%%%%%%%%%%%%%%%%%%%%%%%%%%%%%%%%%
%%%%%%%%%%%%%%%%%%%%%%%%%%%%%%%%%%%%%%%%%%%%%%%%%%%
%--------------------------
\subsection{The Fornax dSph}\label{sec:fornax_density}
\subsubsection{Density profile}
For the Fornax dSph we adopt the observationally-constrained density profile suggested by \citet{Cole2012}. Using the data sets of more than 2000 stars in the Fornax dSph, \citet{Cole2012} applied a Markov Chain Monte Carlo (MCMC) technique to find a family of four dynamical models which best describe the observational data. The models are labelled as large core (LC), weak cusp (WC), intermediate cusp (IC) and steep cusp (SC) and they all have the following mathematical form
\begin{equation}\label{eq:rho_cole}
\rho(r)\propto{\left(\frac{r}{\rs}\right)^{-\gamma_0} \left(1 + \left(\frac{r}{\rs}\right)^{\eta}\right)^{\frac{\gamma_0 - \gamma_\infty}{\eta}} \sech\left(\frac{r}{10\kpc}\right)},
\end{equation}
where $\rs$, $\gamma_0$, $\gamma_\infty$ and $\eta$ are the parameters of the model and their best-fitting values are given in \tabref{tab:cole_params} for different density models.

%--------------------------
\begin{table}
    \centering
	\begin{tabular}{ccccccc}
		\hline
		Model & $M_\infty$ & $\gamma_0$ & $\gamma_\infty$ & $\eta$ & $\rs$ & $A$ \\
		& $(10^8\Msun)$ & & & & $(\kpc)$ & \\
		\hline
		LC & \centering 8.00 & 0.07 & 4.65 & 3.70 & 1.40 & 0.035\\ 
		WC & \centering 1.23 & 0.08 & 4.65 & 2.77 & 0.62 & 0.071\\ 
		IC & \centering 1.51 & 0.13 & 4.24 & 1.37 & 0.55 & 0.190\\ 
		SC & \centering 1.98 & 0.52 & 4.27 & 0.93 & 0.80 & 0.151\\ 
		\hline
	\end{tabular}
    \caption{The adopted density profile models for the Fornax dSph as described by \equref{eq:rho_cole}. \textit{Column designations}: density model (LC: large core, WC: weak cusp, IC: intermediate cusp, and SC: steep cusp), total mass, logarithmic density slope $\left(\gamma \equiv \mathrm{d}\ln(\rho) / \mathrm{d} \ln(r)\right)$ at $r=0$ and $r=\infty$, $\eta$ is a constant and $\rs$ is the scale radius. The values of the parameters are directly brought from Table 2 of \citet{Cole2012} except for the last column, which is the normalization factor (or proportionally constant) and is computed numerically using equations \ref{eq:cole_mass} and \ref{eq:cole_mass_total}.}
    \label{tab:cole_params}
\end{table}

In order to calculate the enclosed mass $M(r)$ within the radius $r$, we integrate over the density function from $\rmin=0$ to $\rmax=r$

\begin{equation}\label{eq:cole_mass}
	M(r) = A {\int_{0}^{r}4\pi r^2\rho \mathrm{d}r},
\end{equation}
where $A$ is the normalization factor. It is computed numerically by calculating the limit of \equref{eq:cole_mass} when $r$ goes to infinity, i.e. 
\begin{equation}\label{eq:cole_mass_total}
    \lim_{r\to\infty}M(r)=M_\infty=\Mtotal
\end{equation}
The left panel of \figref{fig:combined_profiles} shows the mass distribution as a function of radius for different density models of the Fornax dSph  using the parameters given in \tabref{tab:cole_params}. The right panel of \figref{fig:combined_profiles} represents the periodic variation of the tidal radius for different density models of the Fornax dSph while orbiting the MW which is calculated from \equref{eq:r_titdal}.

\subsubsection{Initial conditions of the Fornax dSph}
We obtain the initial conditions of the Fornax dSph, following a similar approach to \secref{sec:MW_dsph_init}. Our calculations yield an orbital eccentricity of $e=0.4$, a pregalactic distance of $\Rpre=65.17\kpc$ and a apogalactic distance of $\Rapo=153.30\kpc$ for the Fornax dSph. 
\par We compute the orbit of 4000 initially-bound GCs distributed within the Fornax dSph for each of density profiles given in \secref{sec:fornax_density}. As an example, \figref{fig:orbits} compares the trajectories for two of these GCs in the SC model at $t=3\Gyr$ and $t=13.6\Gyr$. The figure shows the $Y-Z$ projection of the GC trajectories in the rest frames of the Galaxy and the Fornax dSph. Both GCs are initially bound and remain as such even after $3\Gyrs$ (middle column of the figure). After $13.6\Gyrs$, the second GC which is shown in the bottom row of the figure, reaches distances of $\sim60\kpc$ from the center of the Fornax dSph. Such radii are significantly larger than the tidal radius of Fornax, indicating that the GC has completely escaped from the Fornax dSph and now has an orbit independent of the Fornax dSph in the MW halo.

%%%%%%%%%%%%%%%%%%%%%%%%%%%%%%%%%%%%%%%%%%%%%%%%%%%%%%%%%%%%%%%%%%%%%%%%%%%%%
\section{Results}\label{sec:results}
\begin{figure*}
	\centering
	\includegraphics[width=\textwidth]{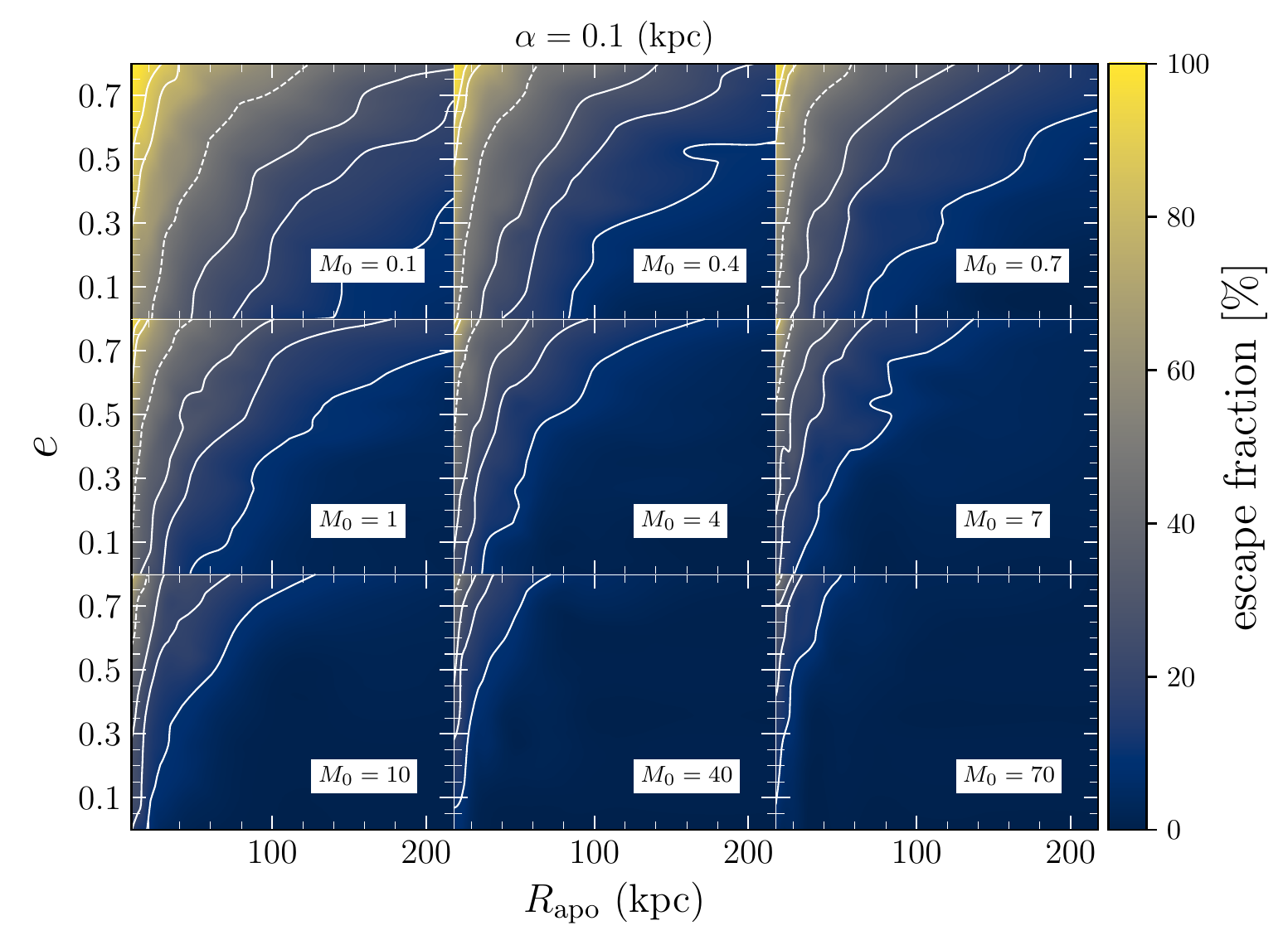}  	
	\caption{The escape fraction of GCs from dSphs, after $13.6\Gyrs$, as a function of the orbital eccentricity and the apogalactic distance. The mass profile for dSphs is a Hernquist model with a scale radius of $\alpha=0.1\kpc$ and different masses which is shown in the white boxes (in units of $10^8\Msun$). The solid and dashed contours correspond to escape fractions of $\{10\%, 20\%, 30\%, 80\%, 90\%\}$ and $\{50\%\}$ respectively.}  
	\label{fig:esc1}
\end{figure*}
\begin{table*}
\begin{tabular}{ccccccccccc}
    \hline
        &  &  &  &  &  &  &  &  &  &   \\
   Galaxy & $M (\rh)$ & $\rh$ & $\Rapo$ & $e$ & $i$ & $M_{1}$ & $\alpha_1$ & $\fesc$ (interp) & $\fesc$ (comp) & $\Tesc$ \\
     & $(10^8 \Msun)$ & ($\kpc$) & ($\kpc$) &  & (deg) & $\left(10^8\Msun\right)$ & $\left(\kpc\right)$ & $\left(\%\right)$ & $\left(\%\right)$ & $\left(\Gyr\right)$ \\
	\hline 
	&  &  &  &  &  &  &  &  &  & \\
    Antlia II & 0.55$^{[1]}$ & 2.920$^{[1]}$ & 136.42$^{+0.50}_{-0.39}$ & 0.68$^{+0.07}_{-0.07}$ & 30.05$^{+8.32}_{-9.96}$ & 1.320 & 1.608 & 91 & ${93}^{+3}_{-4}$ & ${4.3}^{+0.3}_{-0.1}$  \\
    &  &  &  &  &  &  &  &  &  & \\
    CanVen I& 0.19$^{[2]}$ & 0.564$^{[2]}$ & 236.63$^{+24.78}_{-4.18}$ & 0.56$^{+0.21}_{-0.27}$ & 96.64$^{+3.06}_{-9.19}$ & 0.456 & 0.310 &31& ${28}^{+18}_{-15}$ & ${6.8}^{+0.8}_{-0.3}$  \\
    &  &  &  &  &  &  &  &  &  & \\
    Carina I& 0.061$^{[2]}$ & 0.241$^{[2]}$ & 106.84$^{+0.009}_{-0.004}$ & 0.32$^{+0.05}_{-0.05}$ & 82.58$^{+3.40}_{-3.58}$ & 0.146 & 0.132 & 25 & ${25}^{+3}_{-2}$ & ${5.3}^{+0.2}_{-0.4}$  \\
    &  &  &  &  &  &  &  &  &  &   \\
    Draco I& 0.094$^{[2]}$ & 0.196$^{[2]}$ & 83.11$^{+0.41}_{-0.36}$ & 0.50$^{+0.02}_{-0.02}$ & 71.59$^{+2.11}_{-2.19}$& 0.226 & 0.107 & 30 & ${31}^{+3}_{-1}$ & ${4.9}^{+0.1}_{-0.6}$  \\ 
    &  &  &  &  &  &  &  &  &  &   \\
    Fornax I& 0.53$^{[2]}$ & 0.668$^{[2]}$ & 153.30$^{+0.22}_{-0.20}$ & 0.40$^{+0.01}_{-0.01}$ & 108.89$^{+0.51}_{-0.53}$  & 1.274 & 0.367 & 30 & ${27}^{+1}_{-1}$ & ${6.1}^{+0.3}_{-0.1}$  \\
    &  &  &  &  &  &  &  &  &  &   \\
    Leo I& 0.12$^{[2]}$ & 0.246$^{[2]}$ & 337.47$^{+55.15}_{-9.37}$ & 0.79$^{+0.13}_{-0.26}$ & 49.05$^{+74.09}_{-0.50}$ & 0.288 & 0.135 & 18 & ${15}^{+16}_{-10}$ & ${7.0}^{+1.5}_{-0.5}$ \\
    &  &  &  &  &  &  &  &  &  &   \\
    Sculptor I& 0.13$^{[2]}$ & 0.260$^{[2]}$ & 95.73$^{+0.43}_{-0.40}$ & 0.31$^{+0.01}_{-0.01}$ & 87.30$^{+0.12}_{-0.12}$  & 0.312 & 0.143 & 25 & ${23}^{+1}_{-1}$ & ${4.7}^{+0.3}_{-0.1}$  \\ 
    &  &  &  &  &  &  &  &  &  &   \\
    Sextans I& 0.25$^{[2]}$ & 0.682$^{[2]}$ & 110.94$^{+8.30}_{-5.84}$ & 0.22$^{+0.013}_{-0.002}$ & 40.83$^{+0.57}_{-0.29}$ & 0.601 & 0.375 & 39 & ${36}^{+5}_{-2}$ & ${5.2}^{+0.1}_{-0.1}$  \\
    &  &  &  &  &  &  &  &  &  &   \\
    Ursa Major I& 0.26$^{[2]}$ & 0.318$^{[2]}$ & 124.25$^{+85.86}_{-21.97}$ & 0.10$^{+0.24}_{-0.05}$ & 119.12$^{+6.37}_{-7.32}$ & 0.625 & 0.175 &  14& ${12}^{+4}_{-4}$ & ${5.8}^{+1}_{-0.9}$  \\
    &  &  &  &  &  &  &  &  &  &  \\
    Ursa Minor I& 0.15$^{[2]}$ & 0.280$^{[2]}$ & 82.75$^{+0.45}_{-0.35}$ & 0.47$^{+0.05}_{-0.04}$ & 85.60$^{+3.71}_{-3.65}$& 0.360 & 0.154 & 36 & ${37}^{+5}_{-1}$ & ${4.7}^{+0.1}_{-0.4}$  \\ 
    &   &  &  &  &  &  &  &  &  &   \\
    Sgr I& 1.2$^{[2]}$ & 1.550$^{[2]}$ & 35.82$^{+0.55}_{-0.54}$ & 0.41$^{+0.006}_{-0.006}$ & 79.99$^{+0.24}_{-0.24}$    & 2.886 & 0.853 & 82 & ${83}^{+1}_{-1}$ & ${2.5}^{+0.1}_{-0.1}$  \\
    &  &  &  &  &  &  &  &  &  &   \\
    SMC*& - & - & 68.10$^{+6.83}_{-5.06}$ & 0.05$^{+0.04}_{-0.02}$ & 79.17$^{+1.79}_{-1.84}$ & - & - & - & ${17}^{+1}_{-1}$ & ${6.5}^{+0.3}_{-0.4}$  \\
    &  &  &  &  &  &  &  &  &  &   \\
    LMC*& - & - & 102.52$^{+6.25}_{-5.75}$ & 0.36$^{+0.02}_{-0.02}$ & 84.44$^{+1.21}_{-1.26}$ & - & - & - & ${16}^{+1}_{-1}$ & ${7.9}^{+0.1}_{-0.1}$  \\
    &  &  &  &  &  &  &  &  &  &   \\
    \hline 
\end{tabular}
\caption{Average escape time of GCs from 13 massive dSphs of the MW. The density profile of all MW dSphs is the Hernquist model. We assume that the total mass follows the light profile (Model 1). \textit{Column designations}: the name of the dwarf galaxy, enclosed mass at the projected half-light radius in the units of $10^8{\Msun}$, the projected half-light radius, apogalactic distance of the dSph orbit, orbital eccentricity, orbital inclination, mass of Hernquist model, scale length of Hernquist model, interpolated escape fraction assuming $i=0\degr$, escape fraction using the realistic orbit, average escape time of GCs from each dSph. The values of ${\Rapo}$, $e$ and $i$ are calculated using the data from \tabref{tab:MW_dsphs_params}. Projected half-light radii of dSphs are converted to $\alpha_1$ using \equref{eq:r_alpha}. Interpolated escape fractions obtained from \figref{fig:esc1} to \ref{fig:esc3}.  \textit{References}: (1) \citealt{Torrealba2019}, (2) \citealt{Walker2009}}
\label{tab:MW_dsphs}
\end{table*}

\begin{table}
\begin{tabular}{cccc}
    \hline
    Density  &  & Escape Fraction (\%) & \\
    Model & $R\leq\mathrm{R}(0.2\Mtotal)$ & $R\leq\mathrm{R}(0.5\Mtotal)$ & $R\leq\mathrm{R}(0.9\Mtotal)$ \\
    \hline
    LC  & 4 & 6 & 13 \\
    IC  & 6 & 8 & 27 \\
    WC  & 4 & 5 & 13 \\  
    SC  & 6 & 11 & 38 \\
    \hline
\end{tabular}
\caption{The escape fraction of GCs from the Fornax dSph after $13.6\Gyrs$ of evolution for different density Models. The (cumulative) values are calculated for different radii enclosing $20\%$, $50\%$, and $90\%$ of the total mass of the Fornax dSph. These radii are analogous to Lagrange radii defined in self-gravitating systems such as star clusters. The last column gives the overall escape fraction of each model.}
\label{tab:fornax_esc}
\end{table}

\begin{figure*}
	\centering
	\includegraphics[width=0.88\textwidth]{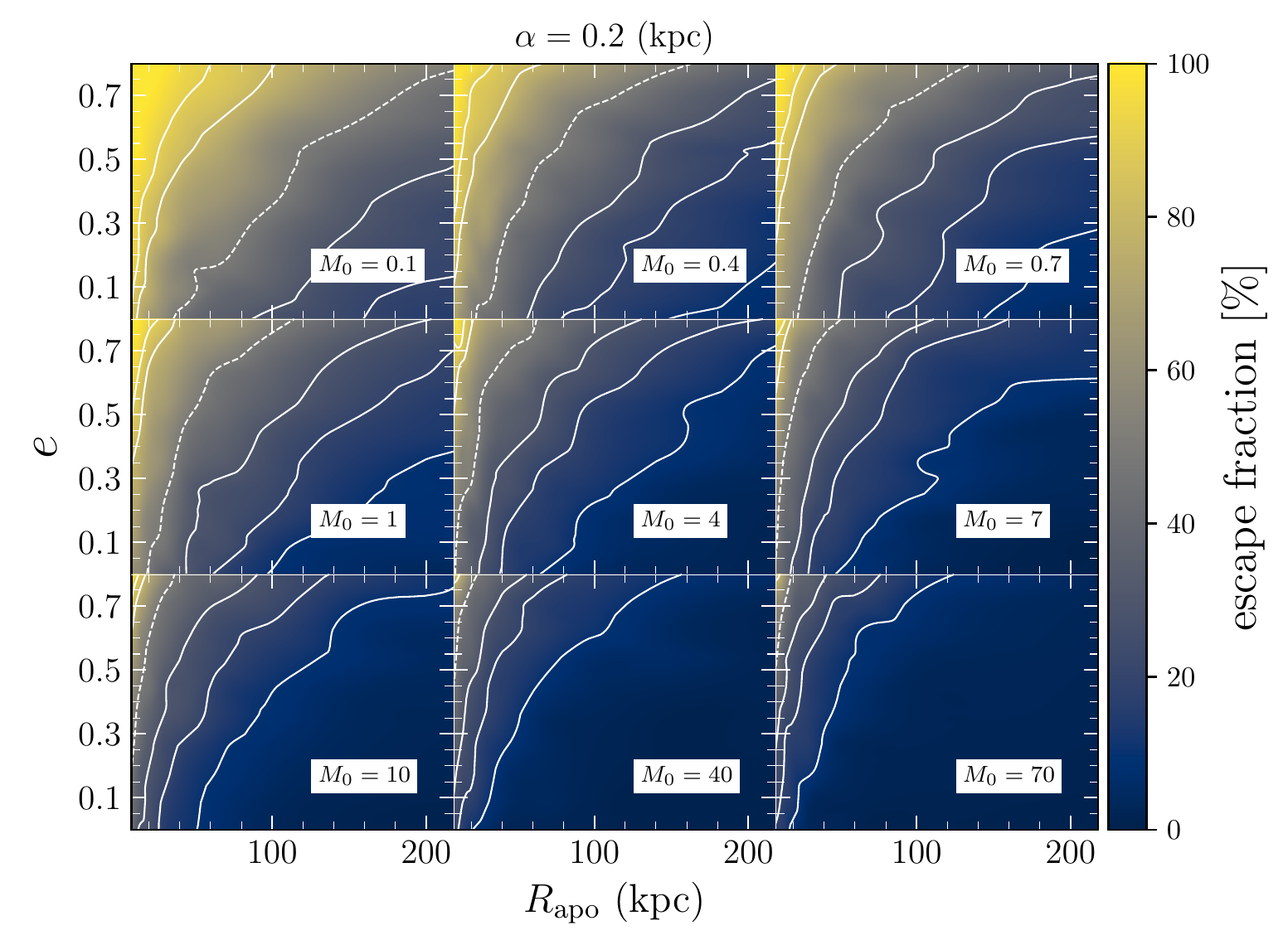}
	\includegraphics[width=0.88\textwidth]{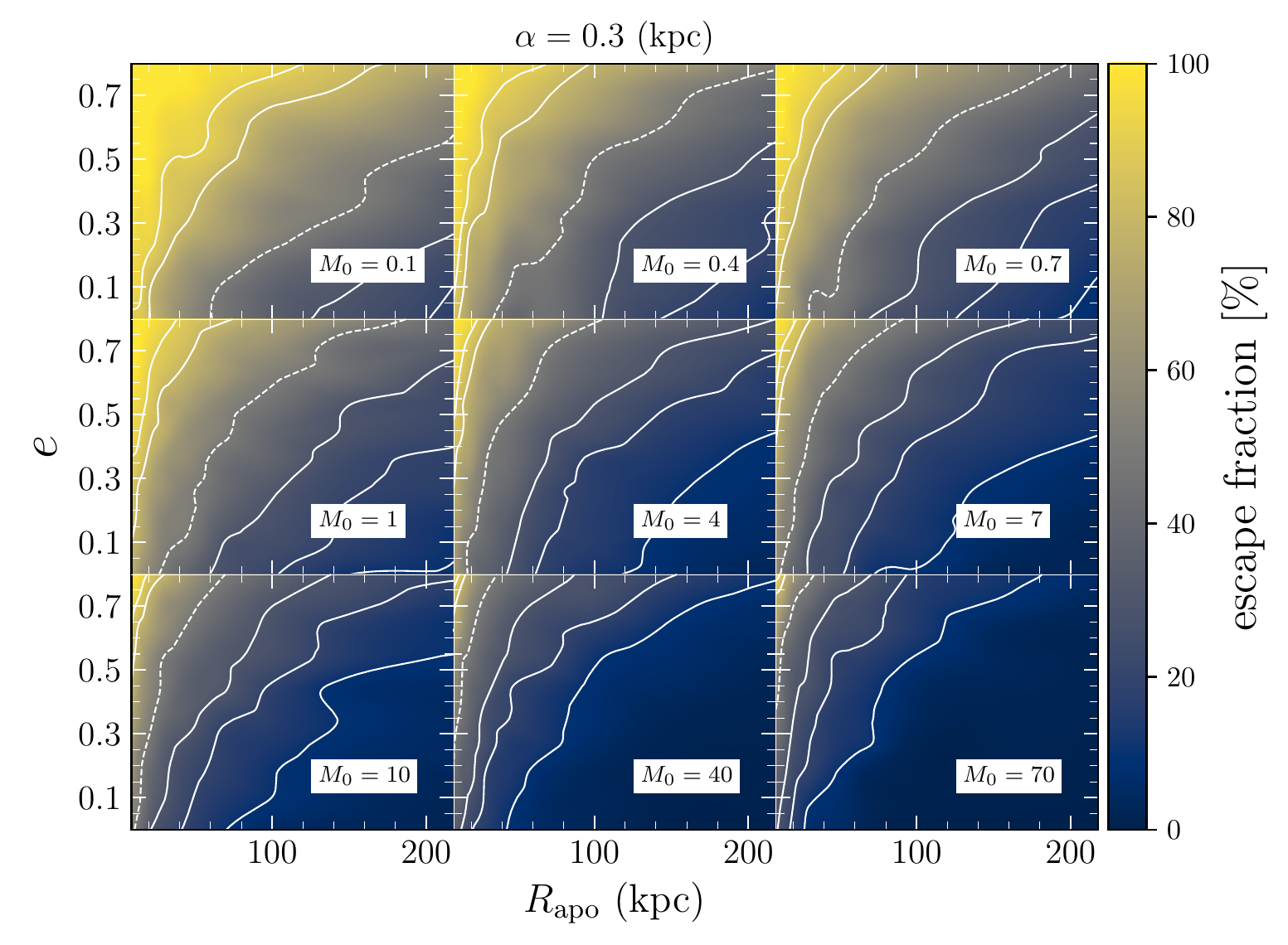}
	\caption{Same as \figref{fig:esc1}, but with the scale radius of $\alpha=0.2\kpc$ (top panel) and $\alpha=0.3\kpc$ (bottom panel). }
	\label{fig:esc2}
\end{figure*}

\begin{figure*}
	\centering
    \includegraphics[width=0.88\textwidth]{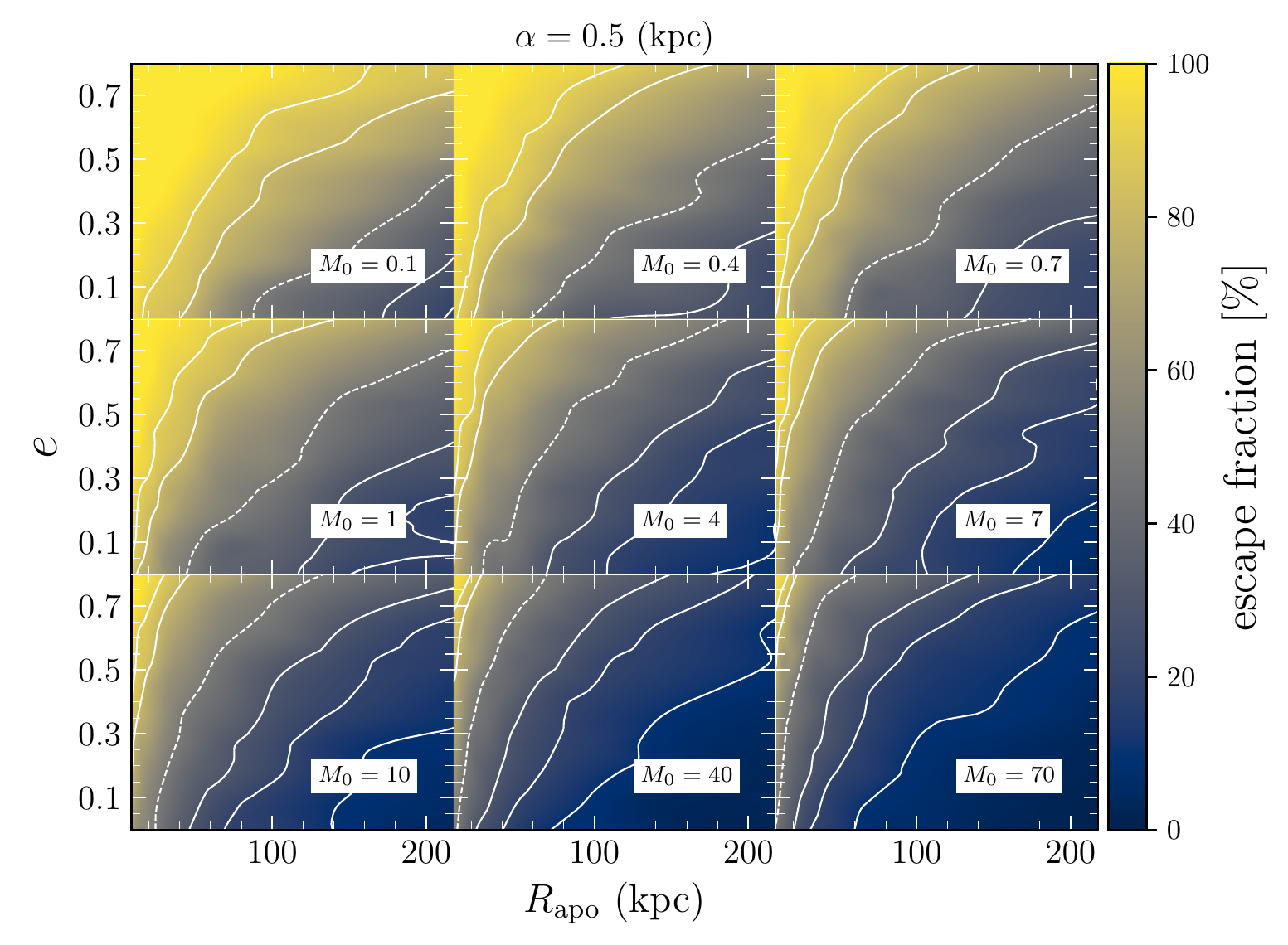}
	\includegraphics[width=0.88\textwidth]{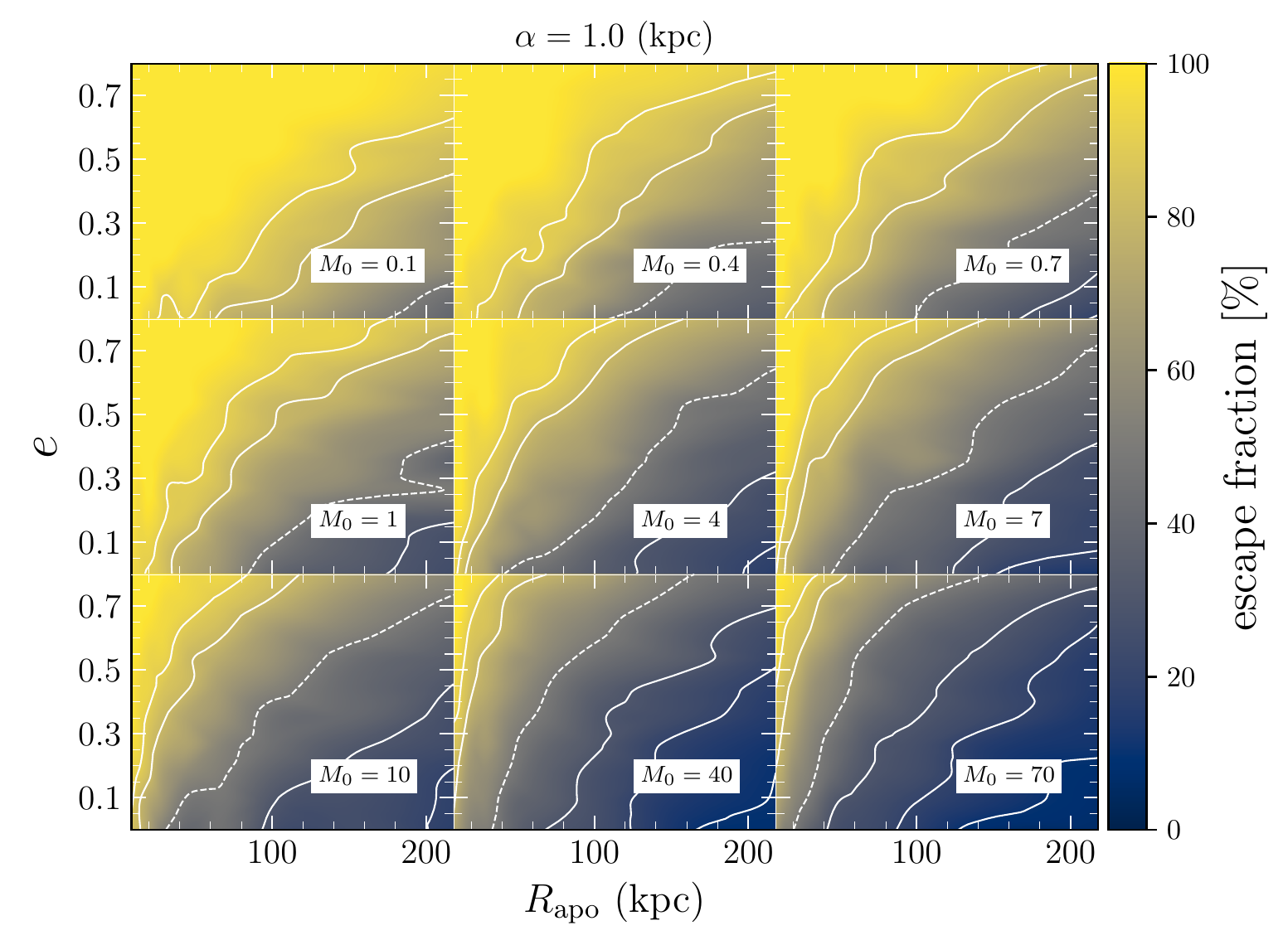}
	\caption{Same as \figref{fig:esc1}, but with the scale radius of $\alpha=0.5\kpc$ (top panel)  and $\alpha=1\kpc$ (bottom panel).}
	\label{fig:esc3}
\end{figure*}

\begin{figure}
	\centering
	\includegraphics[width=1\columnwidth]{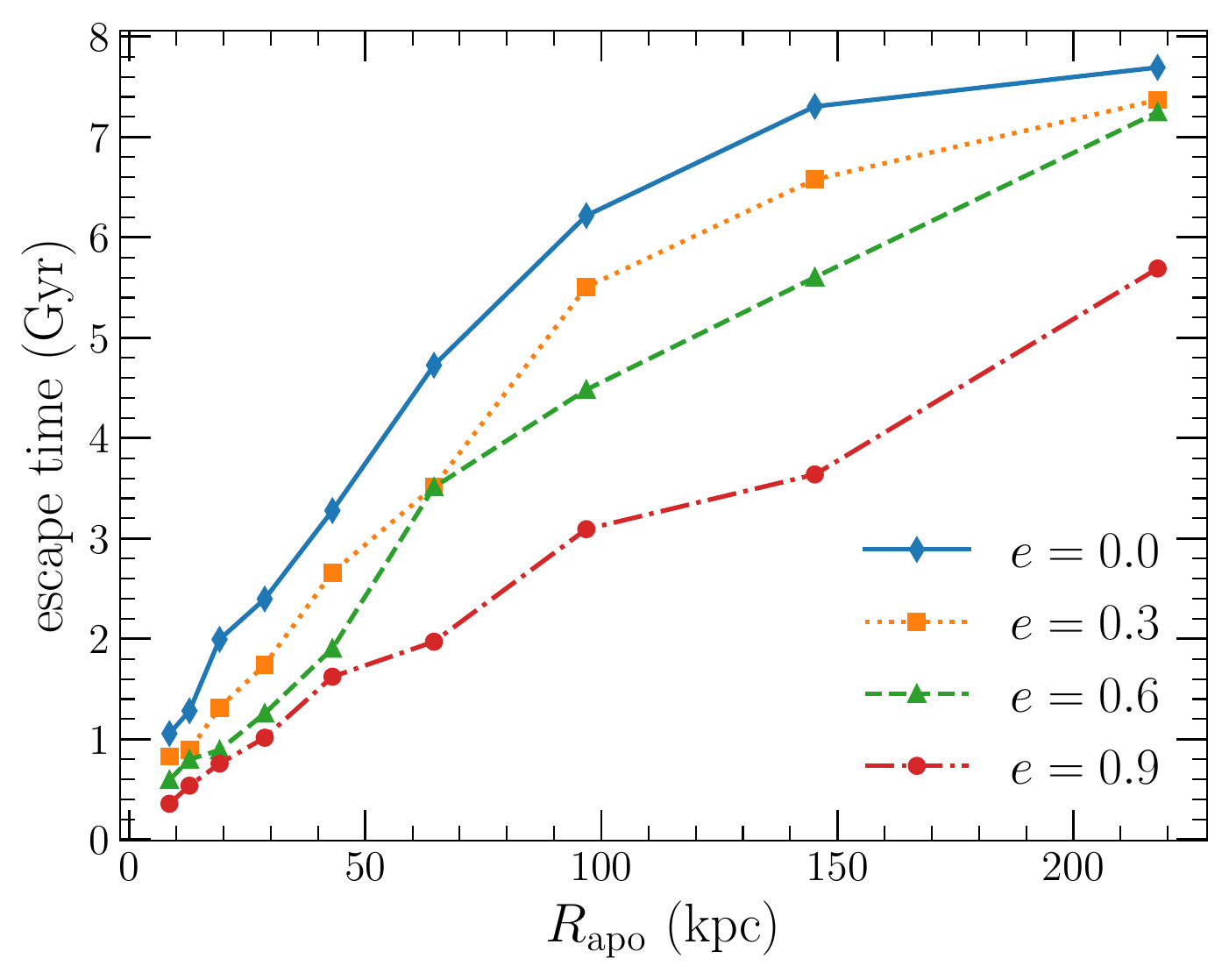}
	\caption{The average escape time of GCs at different orbits of a Fornax-like dSph with SC model on the MW disc plane.}
	\label{fig:escape_time}
\end{figure}

In this section, we present our results for the escape fraction ($\fesc$) and the average escape time ($\Tesc$) of GCs from our modelled dSphs. First, we investigate the general effect of different orbital parameters, scale lengths, and masses of dSphs on the escape fraction of GCs. 

\subsection{The escape fraction of GCs from dSphs: the general case}\label{sec:results_general}
Figures \ref{fig:esc1}, \ref{fig:esc2}, and \ref{fig:esc3} show the escape fraction of GCs from the large sample of dSphs, described in \secref{sec:dsph}, for different values of scale lengths $\alpha$. One can readily identify several trends in the figures. First, there is a significant difference in the escape fraction of GCs from dSphs with different orbital parameters. In particular, dSphs on orbits with smaller apogalactic distances and larger eccentricities demonstrate a more pronounced loss of their GCs as they experience stronger tidal forces. 

\par Second, the escape fraction of GCs is smaller for more compact dSphs, i.e. those with larger masses ($M_0$) and smaller scale radii. This can be explained as follows. The escape speed from an isolated object with a spherically symmetric mass distribution within radius $\alpha$ is $\sqrt{2GM_0/\alpha}$. For an arbitrary mass distribution subject to an external tidal field, the exact mathematical formula for the escape speed is not trivial. However, it still approximately scales with $\sqrt{M_0/\alpha}$ where $\alpha$ is a scale radius in this case. This indicates that the escape speed for more compact and denser dSphs is larger, hence the overall escape fraction has to be smaller. To put this into perspective, for dSphs with $\alpha=0.1\kpc$ and $M_0=7\times10^9\Msun$, the average escape fraction is marginal, i.e. it is less than $20\%$ even under the most extreme tidal conditions ($\Rapo<30\kpc, e>0.7$). On the other side of the spectrum, dSphs with $\alpha=1.0\kpc$ and $M_0=1\times10^7\Msun$ lose at least half of their GCs irrespective of their orbits, and lose more than $90\%$ of their GCs when $\Rapo<100\kpc$ and $e>0.4$. 

\subsection{dSphs of the MW}
\tabref{tab:MW_dsphs} shows our results for the escape fractions of 13 MW dSphs according to Model 1. The results of Model 2 are given in \tabref{tab:MW_dsphs_model2}. We also calculate the average escape time of GCs from dSphs in both density profile models. As it can be seen from the tables, the escape fractions and average escape times for the 8 dSphs in Model 2 closely follow those of Model 1. This implies that one can simply assume a approximate density profile for the dSphs according to Model 1 and obtain escape fractions, which are consistent with those obtained from the Jeans analysis (Model 2).
\par One might be tempted to apply interpolation on figures \ref{fig:esc1} to \ref{fig:esc3} and estimate the escape fractions of GCs in the MW dSphs, given the mass, scale length, and orbital parameters ($a, e$) of the dSph. However, our results are obtained assuming that all dSphs lie on the Galactic disc plane ($i=0\degr$). This is not necessarily the case for the MW dSphs (e.g. \citealt{Pawlowski2013}). Therefore, one needs to verify the applicability of our results for the cases that $i\neq0\degr$. To this end, we compare the escape fractions of GCs from
the MW dSphs using their realistic orbits with the estimated escape fractions obtained by applying interpolation to figures \ref{fig:esc1} to \ref{fig:esc3}. The outcome of this comparison is given in \tabref{tab:MW_dsphs}. One can evidently see that the interpolated and the computed values match very well. This indicates that, under the assumptions of our study, orbital inclination of dSphs, does not play a major role in the average escape fraction of GCs after a Hubble time. One explanation to this could be the following. The mathematical expressions which describe the potential field of the bulge and the halo are both spherically symmetric. Second, in contrast to the Galactic dark-matter halo, the contribution of the Galactic bulge and disc to the overall potential field of the MW asymptotically tends to zero as distance tends to infinity. Therefore, the MW potential field at large radii is mainly governed by a logarithmic dark-matter halo. Both of these, render the orbital inclination only marginally effective in specifying the escape fraction of GCs. Moreover, our criterion for the escape of GCs ($2\times\Rtidal$ of the dSph) is conservative, hence the actual escape fractions could be larger than what we have found.
\par \cite{Massari2019} studied the origin of 151 MW GCs, and concluded that only $40\%$ of MW GCs have an in-situ origin, and that $35\%$ of them can be attributed to known merging processes. They did not find an origin for the rest of the MW GCs. Our results demonstrate that, the escape fraction of GCs from dSphs of the MW is not a negligible value. As a result, it is very likely that a number of MW GCs may have originated from the same dSphs that orbit the MW today. The remainder of this section compares our results for the escape fractions of GCs from the MW dSphs, with those obtained from other studies.
\par We calculated the escape fraction of GCs from the Fornax dSph to be about $25\%$. Since 6 GCs have been observed inside the Fornax today, we expect $\sim2$ GCs to have been escaped from it. In addition, we found that the average escape time of GCs from Fornax is about 6$\Gyr$, implying that the GCs escape process is probably over today. \cite{Mucciarelli} investigated chemical abundances of one of LMC GCs (NGC~2005). They concluded that it may have escaped from the Fornax dSph and then has been captured by the LMC.

\par For the Sgr dSph, there are a number of GCs that are likely to be associated with this dSph, namely Whiting~1, NGC~6715 (M54), Ter~7, Arp~2, Ter~8, Pal~12, NGC~6284, NGC~5053, NGC~4147, NGC~5634, NGC~5824, Pal~2, and NGC~2419 \citep{Law2010,Massari2019,Bellazzini2020}. Among these GCs, M~54, Arp~2, Ter~7 and Ter~8 are still bound to Sgr \citep{Bellazzini2020}. As a result, among the 13 aforementioned GCs, 4 GCs are still bounded to Sgr and the other 9 GCs have escaped. This translates into an escape fraction of $\sim70\%$. In another study, \cite{arakelyan2020} associated 17 GCs with Sgr, of which 5 GCs have a lower probability to be associated with Sgr. Assuming that all 17 GCs are associated with Sgr and 4 of them are bound to it \citep{Bellazzini2020}, the escape fraction of the GCs from Sgr will be $76\%$. These values are in good agreement with our results. In addition, the average escape time was about $3\Gyrs$, which means that the GCs escape process has been long finished.

\par We obtained the escape fractions of LMC and SMC, without considering the effect of dynamical friction, as ${19}^{+1}_{-3}$ and ${18}^{+2}_{-2}$, respectively. Moreover, the average escape time has been obtained as ${5.3}^{+0.1}_{-0.2}$ and ${5.3}^{+0.3}_{-0.4}$, for LMC and SMC respectively. The values of these two variables, considering dynamic friction, are shown in \tabref{tab:MW_dsphs}. As it can be seen, dynamical friction has a marginal effect on the escape fraction over a Hubble time. However, the effect of dynamical friction on the average escape time is not negligible. The average escape time for the LMC and SMC is $\sim7.9\Gyr$ and $\sim6.5\Gyr$ respectively. Given that the initial orbital radii of the LMC and SMC were larger (due to dynamical friction), their GCs did not initially feel a strong tidal field from the MW. As a result, the GCs could escape from the LMC and SMC only after a long time. Today, LMC and SMC have 19 and 8 GCs, respectively \citep{Law2010}. According to the escape fractions that we have obtained, we expect about 4 GCs of the LMC and 2 GCs of the SMC to have already escaped from their host galaxies.

\subsection{Fornax dSph}
As explained earlier in the paper, the density profile and the orbital parameters that we use for the Fornax dSph are both observationally constrained. \tabref{tab:fornax_esc} summarizes our results for Fornax. We calculate the cumulative escape fraction of GCs from $R=0\kpc$ up to three different radii from the centre of the dSph. The radii encompass $20\%$, $50\%$, $90\%$ of the total mass of the Fornax dSph. The density profile of the Fornax, defined in \equref{eq:rho_cole}, asymptotically goes to zero at infinity, i.e. $M_\mathrm{total}=M_\infty$ occurs at $r\to\infty$. As a result, the last column of \tabref{tab:fornax_esc} can be interpreted as the overall escape fraction of GCs for each density model. 
\par One can see that, depending on the density model, the escape fractions of GCs from the outer shells of the Fornax dSph are larger than the inner parts, by a factor of 3 to 6. In particular, the LC and the SC models have the lowest ($13\%$) and the highest ($38\%$) escape fractions, respectively. This is due to the fact that compared to the SC model, the mass in the LC model is distributed over a larger radius (see Fig. \ref{fig:combined_profiles}). As a result, the strength of the potential field falls off more rapidly in the SC model and GCs in the outer shells are only weakly bound to the dSph. Moreover, these values are consistent with what we presented earlier in the section, i.e. according to \tabref{tab:MW_dsphs} the escape fraction of GCs from the Fornax dSph assuming a Hernquist density model, is $30\%$ for an orbit with $i=0\degr$ (by interpolation), and $27\%$ for a realistic orbit. 

\par For comparison, we also calculate the average escape time of GCs from a sample of Fornax-like dSphs (SC model) orbiting the MW on the Galactic disc plane with different orbital eccentricities and apogalactic distances. As it can be seen from \figref{fig:escape_time}, the average escape time of GCs from Fornax-like dSphs exhibits a positive correlation with the Galactocentric distance and a negative correlation with the orbital eccentricity. This is in agreement with what we found in \secref{sec:results_general}, i.e. the escape fraction of GCs increases in stronger tidal fields (larger values of $e$ and smaller values of $\Rapo$). 

\par In addition, unless the orbit of a dSph is extremely eccentric ($e\approx0.9$), all profiles for the average escape time of GCs level off at $t\approx8\Gyr$. This indicates that, regardless of the orbital parameters, the number of GCs escaping from dSphs after $t>8\Gyr$ remains marginal. As a result, dSphs of the MW with a lifetime greater than $8\Gyrs$, are not likely to lose any more GCs. This means that any dSphs that have been orbiting for the same period in the MW must have a current escape rate close to zero.

\subsection{Orbit of runaway GCs}
The bottom row of \figref{fig:orbits} demonstrates a runaway GC which is initially bound to the Fornax dSph. As it can be seen, the GC closely follows the orbit of the Fornax, even after its escape. We checked whether this holds true for other dSphs with different orbits, masses and half-mass radii. We observed that the orbital parameters of a majority of runaway GCs correlate with those of their host dSphs, in agrement with \cite{Law2010,Bellazzini2020, Massari2019}. This provides us with a vital tool using which we can differentiate GCs formed in-situ versus ex-situ. In particular, by an inspection of the distribution of the MW GCs in the position-momentum space, and comparing them with the outcome of our simulations, we can designate the MW GCs that are likely to have an external origin and identify the dSphs they originated from. A thorough analysis of the results of this part is beyond the scope of the present paper, and will be addressed in our future work (Rostami et al., in preparation).

\subsection{Caveats}\label{sec:caveat}
In this section, we highlight a few possible caveats on our analysis.

First, all dSphs are assumed to be in an equilibrium state, i.e. they have a simplified density profile without any tidal interaction with the MW. As a result, the estimated escape fractions correspond to the present-day properties of the dSphs. However, the properties of dSphs prior to any tidal stripping are the most relevant when estimating how many GCs they have lost. In particular, the evolution of dSphs in a triaxial time-dependent potential for the MW should be considered. This will be a subject of future investigations. All escape fractions reported in the present study can be deemed as upper limits, owing to the fact that the initial masses of dSphs were larger prior to any tidal stripping and they have decreased over a Hubble time. However, given the relatively large distances and masses of some dSphs in our analysis, any tidal effect (both in terms of shocks and Virial perturbations) is expected to be negligible, except for dSphs with small orbital radii.

Second, we have omitted the evolution of the Galactic potential field and its consequences on the early- and long-term evolution of dSphs over a Hubble time. In particular, we have assumed a static potential for the MW with parameters corresponding to the present day properties of the MW.
According to the standard paradigm of galaxy formation, galaxies grow through a continuous accretion of gas and a hierarchical merging with smaller galaxies from high redshift to the present day. This implies that the mass and size of galaxies change with time. Assuming a time-dependent Galactic potential affects the orbital history of halo objects and their internal evolution. \cite{Haghi2015} have investigated both of these effects. They showed that in a time-varying potential, the birth of satellite galaxies, at $13\Gyr$ ago, would have occurred at significantly larger Galactocentric distances, compared to the case of a static potential. For example, they found that the orbital period of the LMC around the MW is about $2.7\Gyr$ in the static potential model, while it is $4\Gyr$ in a time-dependent Galactic potential model. As a consequent, assuming a static potential over a Hubble time for the MW, which is often done in studies, leads to an enhancement of mass-loss, i.e. an overestimation of the escape fraction of GCs from dwarf satellites. This is because in the static potential the galaxy mass on average (over time) is larger than that of a time-dependent potential. Therefore, we conclude that the weaker mass-loss of objects evolving in a weaker tidal field in a time-dependent Galactic potential model (i.e. at a larger Galactocentric distance and within a light-mass Galaxy) leads to lower escape fractions. As a result, after a Hubble time of evolution in the framework of a dynamically evolving Galactic potential, we expect to see more GCs to be retained as compared to in dwarfs which evolve in a galaxy with constant mass components.  Investigating the escape fraction within a galaxy which grows with time is the subject of our upcoming study. 

Third, we assumed that the GC distribution follow the total baryonic mass distribution, motivated by \cite{Hudson2018} who obtained the effective radius for the MW GC system to be about $4\kpc$. This assumption however, might be questionable in some galaxies. Any variation in the initial distribution of GCs would affect the estimated escape fractions. For example, \cite{Hudson2018} analysed the radial density profiles of GCs around galaxies in nearby galaxy groups and found that radial density profiles of the GC systems are well fit with a de~Vaucouleurs' profile. They showed that there exists a tight relationship  between the effective radius of the GC system ($R_{\mathrm{e},\mathrm{GCS}}$) and the galaxy light effective radius ($R_{\mathrm{e},\mathrm{light}}$) as $R_{\mathrm{e},\mathrm{GCS}}=3.5R_{\mathrm{e},\mathrm{light}}$. In order to estimate how the escape fractions depend on the initial GCs distribution we have calculated the escape fraction assuming the GC distributions follow the dark matter distribution (i.e., a wider spatial distribution).  Since the dark matter halo is distributed over much farther distances compared to the stellar distribution, the escape fraction increases by a factor of about 3. This shows that the escape fractions are sensitive to the initial GCs distribution.

%--------------------------------------------------------------------------------------------------------

\section{Conclusion}\label{sec:conclusion}
According to the outcomes of cosmological simulations, galaxies are formed through a mixture of accretion and in-situ star formation, hence they may contain both in-situ and accreted GCs. As fossil records of the chemical and dynamical evolution of galaxies, the orbital history of GCs are important tools for understanding how in-situ and accreted GCs differ. 

In order to estimate the fraction of the MW GCs that may have originated in dwarf galaxies and later accreted on to the MW, we carried out a grid of runs for a large sample of three-body systems consisting of a GC that forms in dSphs, orbiting the MW. GCs move in the combined gravitational potential field of their host dSphs and the Galactic field. In the first part of our study, we assumed a general density profile (i.e. Hernquist model), but with different masses and half-mass radii (scale lengths). Our modelled dSphs orbit the MW in the Galactic disc plane but they have different orbital parameters (apogalactic distance and eccentricity). We followed the trajectories of this large ensemble of dSphs and their GCs in the halo of the MW for a period of 13.6 billion years. We then obtained the fraction of GCs that leave their dSphs after a Hubble time.

Next, we computed the escape fraction of GCs from 13 of the most massive dSphs of the MW, using their realistic orbits. To achieve this, we utilised the observed proper motions, coordinates and heliocentric distances of the MW dSphs obtained from \textit{Gaia} DR2. Furthermore, to see the effect of different density profiles on the escape fractions of GCs, we focused on the Fornax dSph as an example. We adopted a set of observationally-constrained density profiles for Fornax, which cover a wide range of possible density/potential profiles for this dSph, ranging from models with a large core down to a steep cusp. Finally, to investigate the effect of orbital parameters on average escape time of GCs from dSphs, we repeated our calculations for a large number of Fornax-like dSphs, orbiting on the galactic disc plane with different orbital eccentricities and apogalactic distances.

In total, we have more than 1.8 million three-body systems (GC, dSph, MW), for which we numerically solve the equations of motion using a 10th order Runge-Kutta integrator. The main outcomes of our study can be summarized as follows.
\begin{itemize}

\item We observed that while the dSphs gradually spiral towards the Galaxy, they will lose some of their GCs which will contribute to the Galactic GCs in the halo. The escape fraction of GCs strongly depends on the apogalactic distance and the orbital eccentricity. In particular, it anti-correlates both with the mass and the orbital apogalactic distance of the dSphs. On the other hand, it increases with orbital eccentricity of the dSphs (figures \ref{fig:esc1} to \ref{fig:esc3}).

\item Orbital inclination of the dSphs also has an effect on the escape fraction of GCs, albeit to a much weaker extent especially as the distance of dSphs increase (\tabref{tab:MW_dsphs}). This can be attributed to the isotropy of the assumed profiles for the Galactic bulge and dark matter halo. Moreover, the effect of the halo on the overall potential field of the MW dominates over the Galactic disc and bulge at larger radii. 

\item In general, the escape fraction of GCs from MW dSphs are not negligible. For most dSphs this value was at least $\sim20\%$ and for two of them it was above $80\%$. Therefore, it is very likely that a number of MW GCs originated from its dSphs. Given the number of GCs observed within the Fornax, Sgr, SMC and LMC, and the escape fractions we obtained for them, we expect at least 2 GCs to have escaped from Fornax, 2 and 4 GCs have escaped from SMC and LMC respectively, and about 14 GCs have escaped from Sgr. Our results for Sgr is in agreement with previous studies on the association of some MW GCs with Sgr.

\item The escape fraction of GCs from the Fornax dSph ranges from $13\%$ to $38\%$ depending on the adopted initial density distribution (\tabref{tab:fornax_esc}). In particular, among all the models that we considered for the Fornax, models with large cores and steep cusps have the lowest and the highest escape fractions. 

\item The average escape time of GCs from the sample of Fornax-like dSphs increases with the apogalactic distance and decreases with the orbital eccentricity (\figref{fig:escape_time}). It reaches a plateau at $t\approx8\Gyrs$ which means that dSphs of the MW with a lifetime greater than $8\Gyrs$, are not likely to lose any more GCs. This holds true for eccentricities smaller than 0.9, irrespective of the apogalactic distance.

\item The computed escape fraction of GCs from the Fornax (using a realistic orbit and density profile) as well as other MW dSphs (using realistic orbits but a generic density model), show a remarkable agreement with the estimated escape fraction of GCs; which is derived from interpolation, given ($M$, $r_h$, $\Rapo$, $e$) of the dSphs only (\tabref{tab:MW_dsphs}). 

\item We observed that final orbit of runaway GCs in the MW halo does not differ significantly from the orbit of their host dSphs. We will further elaborate upon this in a future work (Rostami et al., in preparation).

\item To calculate the long-term orbital evolution of the GCs initially distributed in dSphs, we consider a static gravitational potential both for the MW and the dSphs. This can be considered as a caveat of our study. \Nbody simulations of tidal stripping of the GCs from the dwarf galaxies in the time-varying MW potential can give us a more detailed view, and enable us to draw a more robust comparison with the observed GC population of the MW. In particular, our results can be complemented by performing more detailed simulations using codes such as \textsc{SUPERBOX} \citep{Fellhauer2000}. However, a major advantage of our approach over such methods, is its efficiency to comprehensively explore a large multi-dimensional parameter-space over a short period of time.

\end{itemize}

\section*{Data availability}
The data underlying this article are available in the article.

% Bibliography
%-----------------------------------------------------------------
\bibliographystyle{mnras}
\bibliography{article} 

%%%%%%%%%%%%%%%%% APPENDICES %%%%%%%%%%%%%%%%%%%%%
\appendix
\section{Orbital Parameters of the MW dSphs}
\begin{figure*}
	\centering
	\includegraphics[width=0.95\textwidth]{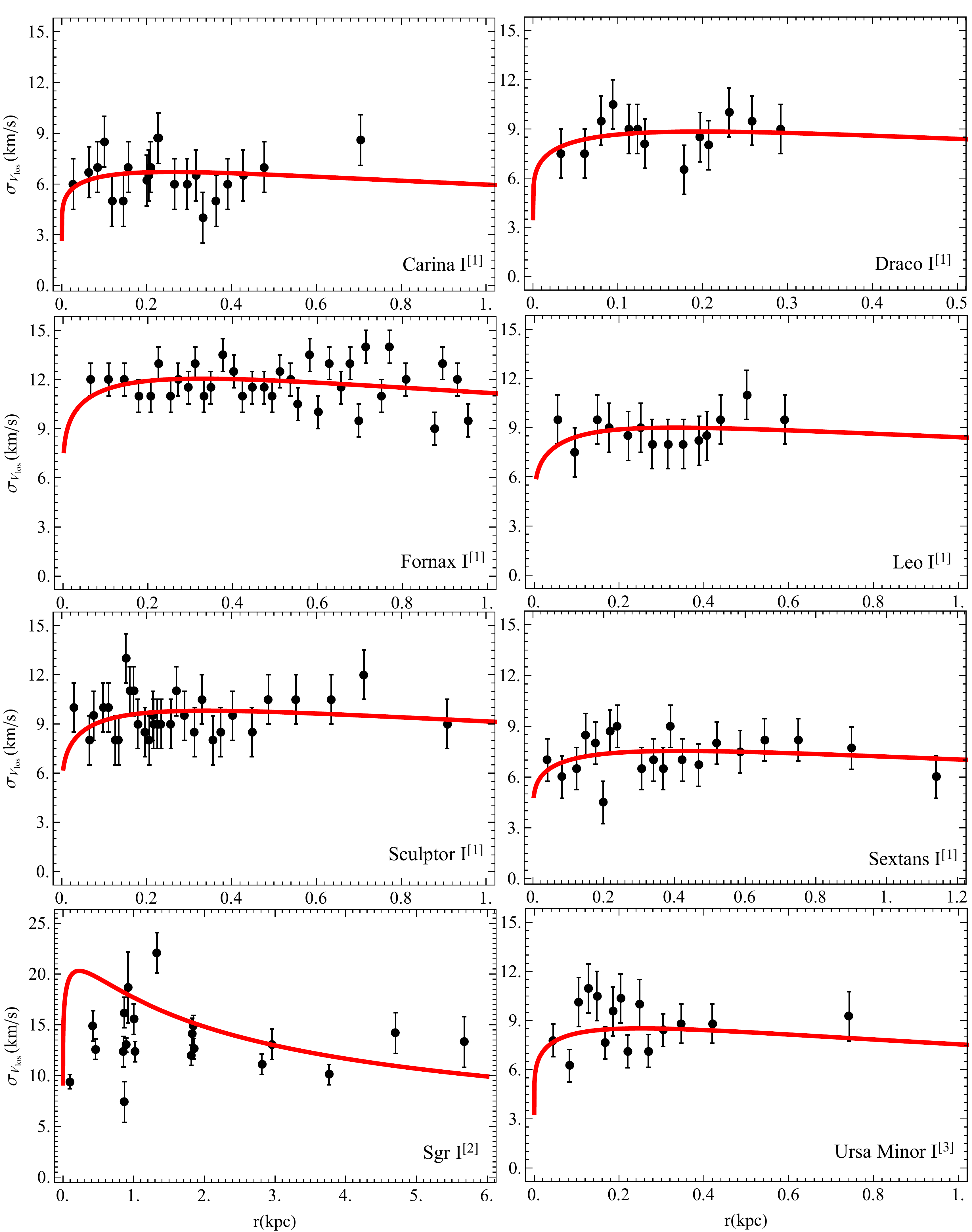}	
	\caption{The line-of-sight velocity dispersion as a function of projected radial distance for 8 bright MW dSphs. The red lines show the best-fitting
solutions obtained by performing a full Jeans analysis. We solved the Jeans equation assuming isotropy for the distributions. \textit{References}: (1) \citealt{angus2008}, (2) \citealt{lokas2010}, (3) \citealt{Walker2009}}  
	\label{fig:vel_dis}
\end{figure*}

\begin{table*}
\begin{tabular}{ccccccc}
    \hline
        &  &  &  &  &  &  \\
   Galaxy& $M_2$ & $\alpha_2$ & $M^{\prime[1]}_{2}$ & $\mathrm{\alpha}^{\prime}_{2}$ & $\fesc$ (comp) & $\Tesc$\\
     & $
     \left(10^8\Msun\right)$ & $\left(\kpc\right)$ & $\left(10^6\Msun\right)$ & $\left(\kpc\right)$ & $\left(\%\right)$ & $\left(\Gyr\right)$ \\
	\hline 
	&  &  &  &  &  & \\
    Carina I& 1.177 & 1.120 & 0.38 & 0.132 & ${27}^{+3}_{-6}$ & ${5.5}^{+0.5}_{-0.4}$\\ 
    &  &  &  &  &  &  \\
    Draco I& 1.577 & 0.870 & 0.29 & 0.107 & ${33}^{+5}_{-2}$ & ${5.1}^{+0.2}_{-0.3}$ \\ 
    &  &  &  &  &  &  \\
    Fornax I& 5.003 & 1.700 & 20 & 0.367 & ${26}^{+3}_{-3}$ & ${6.1}^{+0.4}_{-0.1}$  \\
    &  &  &  &  &  &   \\
    Leo I& 3.158 & 1.879 & 5.5 & 0.135 & ${17}^{+19}_{-13}$ & ${7.0}^{+1.6}_{-0.2}$  \\
    &  &  &  &  &  &   \\
    Sculptor I& 3.503 & 1.641 & 2.3 & 0.143 & ${23}^{+1}_{-1}$ & ${5.5}^{+0.2}_{-0.4}$\\ 
    &  &  &  &  &  &  \\
    Sextans I& 2.422 & 1.841 & 0.44 & 0.375 & ${37}^{+5}_{-3}$ & ${5.9}^{+0.1}_{-0.5}$  \\
    &  &  &  &  &  &  \\
    Ursa Minor I& 1.869 & 1.112 & 0.29 & 0.154 & ${43}^{+9}_{-5}$ & ${4.9}^{+0.1}_{-0.5}$\\ 
    &  &  &  &  &  & \\
    Sgr I& 9.496 & 1.027 & 21 & 0.853 & ${77}^{+1}_{-1}$ & ${2.7}^{+0.1}_{-0.1}$ \\
    &  &  &  &  &  &  \\
    \hline 
\end{tabular}
\caption{Escape fraction and average escape time of GCs from 8 dSphs of the MW (Model 2). The density profile of dSphs (including baryonic matter and halo) follows the Hernquist model. For the baryonic part, the mass ($M^{\prime}_{2}$) is the stellar mass and the scale length ($\mathrm{\alpha}^{\prime}_{2}$) obtained from projected half-light radii. Mass ($M_2$) and scale length ($\alpha_2$) of the halo, is obtained from best-fitting solutions of the Jeans equation. \textit{Column designations}: the name of the dwarf galaxy, mass of the Hernquist profile of the halo, scale length of the halo profile, mass of the Hernquist profile for baryonic core, scale length of the core, escape fraction of GCs, average escape time of GCs from each dSph. \textit{Reference}: (1) \citealt{Alan2012}}
\label{tab:MW_dsphs_model2}
\end{table*}

\begin{table*}
\begin{tabular}{cccccccccccc}
    \hline
    & & & & & & \\
	Dwarf Galaxy & $\alpha$ & $\delta$ & $D_{\odot}$ & $\mu_{\alpha}\cos{\delta}$ & $\mu_{\delta}$ & $\VLOS$\\
	 & (deg) & (deg) & $(\kpc)$ & $(\mathrm{mas\,yr}^{-1})$ & $(\mathrm{mas\,yr}^{-1})$ & $(\kms)$ \\
	\hline \\
    Antlia II     & $143.88^{[5]}$  & $-36.76^{[5]}$  & $132^{[5]}$   & $-0.095\pm0.018^{[5]}$ & $ 0.058\pm0.024^{[5]}$ & $  290.7\pm0.5^{[5]}$ \\ \\ 
    CanVen I      & $202.01^{[6]}$  & $ 33.55^{[6]}$  & $218^{[2]}$   & $-0.159\pm0.094^{[1]}$ & $-0.067\pm0.054^{[1]}$ & $  30.9\pm0.6^{[1]}$  \\ \\  
    Carina I      & $100.40^{[3]}$  & $-50.96^{[3]}$  & $105.2^{[7]}$ & $ 0.485\pm0.017^{[1]}$ & $ 0.131\pm0.016^{[1]}$ & $ 229.1\pm0.1^{[1]}$ \\ \\ 
    Draco I       & $260.05^{[3]}$  & $ 57.91^{[3]}$  & $ 75.9^{[7]}$ & $-0.012\pm0.013^{[1]}$ & $-0.158\pm0.015^{[1]}$ & $-291\pm0.1^{[1]}$  \\ \\ 
    Fornax I      & $ 39.99^{[3]}$  & $-34.44^{[3]}$  & $147.2^{[7]}$ & $ 0.374\pm0.004^{[1]}$ & $-0.401\pm0.005^{[1]}$ & $  55.3\pm0.3^{[1]}$ \\ \\ 
    Leo I         & $152.11^{[3]}$  & $ 12.30^{[3]}$  & $253.5^{[7]}$ & $-0.086\pm0.059^{[1]}$ & $-0.128\pm0.062^{[1]}$ & $ 282.5\pm0.5^{[1]}$ \\ \\  
    Sculptor I    & $ 15.03^{[3]}$  & $-33.70^{[3]}$  & $ 85.9^{[7]}$ & $ 0.084\pm0.006^{[1]}$ & $-0.133\pm0.006^{[1]}$ & $ 111.4\pm0.1^{[1]}$ \\ \\
    Sextans I     & $153.26^{[3]}$  & $ -1.61^{[3]}$  & $ 85.9^{[7]}$ & $-0.438\pm0.028^{[1]}$ & $ 0.055\pm0.028^{[1]}$ & $ 224.2\pm0.1^{[1]}$ \\ \\
    Ursa Major I  & $158.72^{[4]}$  & $ 51.92^{[4]}$  & $ 97.3^{[4]}$ & $-0.683\pm0.094^{[1]}$ & $-0.72\pm0.13^{[1]}$  & $ -55.3\pm1.4^{[1]}$ \\ \\
    Ursa Minor I  & $227.28^{[3]}$  & $ 67.22^{[3]}$  & $ 75.9^{[7]}$ & $-0.184\pm0.026^{[1]}$ & $ 0.082\pm0.023^{[1]}$ & $-246.9\pm0.1^{[1]}$  \\ \\
    Sgr I         & $283.83^{[3]}$  & $-30.54^{[3]}$  & $ 26.3^{[7]}$ & $-2.736\pm0.009^{[1]}$ & $-1.357\pm0.008^{[1]}$ & $ 140\pm 2^{[1]}$ \\ \\
    SMC           & $12.80^{[3]}$ & $-73.15^{[3]}$ & $ 64^{[7]}$   & $ 0.797\pm0.03^{[3]}$ & $-1.220\pm0.03^{[3]}$ & $ 145.6\pm0.6^{[3]}$ \\ \\
    LMC           & $81.28^{[3]}$ & $-69.78^{[3]}$ & $ 50.6^{[7]}$ & $ 1.850\pm0.03^{[3]}$ & $ 0.234\pm0.03^{[3]}$ & $ 262.2\pm3.4^{[3]}$  \\ \\ 
    \hline
    
\end{tabular}    
 \caption{Astrometric data of 13 of the most massive dSphs of the MW. For all rows, columns 2 and 3 represent the equatorial coordinates $(\alpha, \delta)$. The heliocentric distance of each dSph is denoted $D_{\sun}$. Columns 5 and 6 show the components of proper motions in the direction of right ascension ($\mathrm{\mu}_{\alpha}\cos{\delta}$) and declination ($\mathrm{\mu}_{\delta}$), respectively. The last column represents the line-of-sight velocity ($V_\mathrm{LOS}$). \textit{References}: (1) \citet{Fritz2018}, (2) \citet{Alan2012}, (3) \citet{Helmi2018}, (4) \citet{Simon2018}, (5) \citet{Torrealba2019},(6) \citet{Alan2020}, (7) \citet{Pawlowski2013}}
\label{tab:MW_dsphs_params}
\end{table*}

%%%%%%%%%%%%%%%%%%%%%%%%%%%%%%%%%%%%%%%%%%%%%%%%%%%%%%%%%%%%%%%%%%%%%%5
\bsp
\label{lastpage}
\end{document}